\documentclass[sigconf,screen]{acmart}

\usepackage{xcolor} 
\usepackage[nomessages]{fp} 
\usepackage{booktabs} 
\usepackage{multirow} 
\usepackage{url}
\usepackage{framed}
\usepackage{enumitem}
\usepackage{array}
\usepackage{tabularx}
\usepackage{balance}
\usepackage{amsmath} 
\usepackage{siunitx} 
\usepackage{subcaption}
\usepackage{adjustbox}
\usepackage[flushleft]{threeparttable}
\usepackage{setspace} 
\usepackage{caption} 
\usepackage[font=itshape]{quoting}

\definecolor{ForestGreen}{rgb}{0.13, 0.55, 0.13}
\definecolor{tomBlue}{HTML}{0196CE}

\usepackage{soul,color}
\soulregister\cite7
\soulregister\ref7
\soulregister\pageref7
\setstcolor{red}
\setul{}{.2ex}

\newcolumntype{L}[1]{>{\raggedright\let\newline\\\arraybackslash\hspace{0pt}}m{#1}}
\newcolumntype{C}[1]{>{\centering\let\newline\\\arraybackslash\hspace{0pt}}m{#1}}
\newcolumntype{R}[1]{>{\raggedleft\let\newline\\\arraybackslash\hspace{0pt}}m{#1}}

\usepackage{xspace}

\newcommand{\cem}{\texttt{CEM}$^{\mbox{\tiny{ORD}}}$\xspace}

\newcommand{\covid}{COVID-19\xspace}

\newcommand{\six}{C$_{6}$\xspace}
\newcommand{\three}{C$_{3}$\xspace}
\newcommand{\two}{C$_{2}$\xspace}

\newcommand{\expertsix}{E$_{6}$\xspace}
\newcommand{\expertthree}{E$_{3}$\xspace}
\newcommand{\experttwo}{E$_{2}$\xspace}

\newcommand{\politifact}{\texttt{PolitiFact}\xspace}
\newcommand{\abc}{\texttt{ABC}\xspace}

\newcommand{\politifactpantsfire}{\texttt{pants-on-fire}\xspace}
\newcommand{\politifactfalse}{\texttt{false}\xspace}
\newcommand{\politifactmostlyfalse}{\texttt{mostly-false}\xspace}
\newcommand{\politifacthalftrue}{\texttt{half-true}\xspace}
\newcommand{\politifactmostlytrue}{\texttt{mostly-true}\xspace}
\newcommand{\politifacttrue}{\texttt{true}\xspace}

\newcommand{\politifactthreebinszero}{\texttt{01}\xspace}
\newcommand{\politifactthreebinsone}{\texttt{23}\xspace}
\newcommand{\politifactthreebinstwo}{\texttt{45}\xspace}

\newcommand{\politifacttwoebinszero}{\texttt{012}\xspace}
\newcommand{\politifacttwobinsone}{\texttt{234}\xspace}

\newcommand{\high}{\texttt{High}\xspace}
\newcommand{\low}{\texttt{Low}\xspace}

\newcommand{\myparagraph}[1]{\vspace{0.5\baselineskip}\noindent{\textit{#1}.}~}

\AtBeginDocument{%
  \providecommand\BibTeX{{%
    \normalfont B\kern-0.5em{\scshape i\kern-0.25em b}\kern-0.8em\TeX}}}

\copyrightyear{2020}
\acmYear{2020}
\setcopyright{acmlicensed}
\acmConference[CIKM '20] {The 29th ACM International Conference on Information and Knowledge Management}{October 19--23, 2020}{Virtual Event, Ireland}
\acmBooktitle{The 29th ACM International Conference on Information and Knowledge Management (CIKM '20), October 19--23, 2020, Virtual Event, Ireland}
\acmPrice{15.00}
\acmDOI{10.1145/3340531.3412048}
\acmISBN{978-1-4503-6859-9/20/10}

\author{Kevin Roitero}
\email{roitero.kevin@spes.uniud.it}
\affiliation{
\institution{University of Udine}
\city{Udine}
\country{Italy}
}

\author{Michael Soprano}
\email{soprano.michael@spes.uniud.it}
\affiliation{
\institution{University of Udine}
\city{Udine}
\country{Italy}
}

\author{Beatrice Portelli}
\email{portelli.beatrice@spes.uniud.it}
\affiliation{
\institution{University of Udine}
\city{Udine}
\country{Italy}
}

\author{Damiano Spina}
\email{damiano.spina@rmit.edu.au}
\affiliation{
\institution{RMIT University}
\city{Melbourne}
\country{Australia}
}

\author{Vincenzo Della Mea}
\email{vincenzo.dellamea@uniud.it}
\affiliation{
\institution{University of Udine}
\city{Udine}
\country{Italy}
}

\author{Giuseppe Serra}
\email{giuseppe.serra@uniud.it}
\affiliation{
\institution{University of Udine}
\city{Udine}
\country{Italy}
}

\author{Stefano Mizzaro}
\email{mizzaro@uniud.it}
\affiliation{
\institution{University of Udine}
\city{Udine}
\country{Italy}
}

\author{Gianluca Demartini}
\email{demartini@acm.org}
\affiliation{
\institution{The University of Queensland, Australia}
}

\usepackage{comment}

\begin{document}
\fancyhead{}

\title[The COVID-19 Infodemic: Can the Crowd Judge Recent Misinformation Objectively?]
{The COVID-19 Infodemic: Can the Crowd Judge\\Recent Misinformation Objectively?}

\begin{abstract}
Misinformation is an ever increasing problem that is difficult to solve for the research community and has a negative impact on the society at large. Very recently,
the problem has been addressed with a crowdsourcing-based approach to scale up labeling efforts: to assess the truthfulness of a statement, instead of relying on a few experts, a crowd of (non-expert) judges is exploited.
We follow the same approach 
to study 
whether crowdsourcing is an effective and reliable method to assess 
statements truthfulness 
during a pandemic. We specifically target statements related to the \covid health emergency, that is still ongoing at the time of the study and has arguably caused an increase of the amount of misinformation that is spreading online (a phenomenon for which the term ``infodemic'' has been used). By doing so, we are able to address (mis)information that is both related to a sensitive and personal issue like health and very recent as compared to when the judgment is done: two issues that have not been analyzed in related work.  

In our experiment, crowd workers are asked to assess the truthfulness of statements, as well as to provide evidence for the assessments 
as
a URL 
and a text justification.  
Besides showing that the crowd is able to accurately judge the truthfulness of the statements, we also report results on many different aspects, including: agreement among workers, the effect of different aggregation functions, of scales transformations, and of workers background / bias. We also analyze  workers behavior, in terms of queries submitted, URLs found / selected, text justifications, and other behavioral data like clicks and mouse actions collected by means of an ad hoc logger. 
\end{abstract}


\maketitle

\section{Introduction}\label{sec:intro}

\begin{quoting}[leftmargin=4mm]
\noindent
``We're concerned about the levels of rumours and misinformation that are hampering the response. [\ldots] we're not just fighting an epidemic; we're fighting an infodemic.
Fake news spreads faster and more easily than this virus, and is just as dangerous. That's why we're also working with search and media companies like Facebook, Google, Pinterest, Tencent, Twitter, TikTok, YouTube and others to counter the spread of rumours and misinformation. 
We call on all governments, companies and news organizations to work with us to sound the appropriate level of alarm, without fanning the flames of hysteria.''
\end{quoting}
These are the alarming words used by Dr.~Tedros Adhanom Ghebreyesus, the WHO (World Health Organization) Director General during his speech at the Munich Security Conference on 15 February 2020.\footnote{\url{https://www.who.int/dg/speeches/detail/munich-security-conference}} 
It is telling that the WHO Director General chooses to target explicitly  misinformation related problems. 

Indeed, during the still ongoing \covid health emergency, all of us have experienced mis- and dis-information.  The research community has focused on several \covid related issues \cite{bullock2020mapping}, ranging from machine learning systems aiming to classify statements and claims on the basis of their truthfulness \cite{pennycook2020fighting}, search engines tailored to the \covid related literature, as in the ongoing  TREC-COVID Challenge\footnote{\url{https://ir.nist.gov/covidSubmit/}} \cite{TREC-COVID:JAMIA:2020}, topic-specific workshops like the  NLP COVID workshop at ACL'20,\footnote{\url{https://www.nlpcovid19workshop.org/}} and evaluation initiatives like the 
TREC Health Misinformation Track 2020.\footnote{\url{https://trec-health-misinfo.github.io/}}
More than the academic research community, commercial social media platforms also have looked at this issue.\footnote{\url{https://www.forbes.com/sites/bernardmarr/2020/03/27/finding-the-truth-about-covid-19-how-facebook-twitter-and-instagram-are-tackling-fake-news/} and \url{https://spectrum.ieee.org/view-from-the-valley/artificial-intelligence/machine-learning/how-facebook-is-using-ai-to-fight-covid19-misinformation}}
Among all the approaches, in some very recent work, \citet{RDSM:2018}, 
\citet{labarbera2020crowdsourcing}, \citet{SIGIR:2020} 
have studied if crowdsourcing can be used to identify misinformation. 
As it is well known, \emph{crowdsourcing} means to outsource to a large mass of unknown people (the ``crowd''), by means of an open call, a task that is usually performed by a few experts. That recent work \cite{RDSM:2018,labarbera2020crowdsourcing,SIGIR:2020} specifically crowdsource the task of misinformation identification, or rather assessment of the truthfulness of statements made by public figures (e.g., politicians), usually on political, economical, and societal issues.
That the crowd is able to identify misinformation might sound implausible at first---isn't the crowd the very mean to spread misinformation? However, on the basis of recent research \cite{RDSM:2018,labarbera2020crowdsourcing,SIGIR:2020}, it appears that the crowd can provide high quality truthfulness labels, 
provided that adequate countermeasures and quality assurance techniques are employed.

In this paper we address the very same problem, but focusing on statements about \covid. This is motivated by several reasons. 
First, \covid is of course a hot topic but, although there is a great amount of  researchers efforts worldwide  devoted to its study, there is no study yet using crowdsourcing to assess truthfulness of \covid related statements. To the best of our knowledge, we are the first to report on crowd assessment of \covid related misinformation.
Second, the health domain is particularly sensitive, 
so it is interesting to understand if the crowdsourcing approach is adequate also in such a particular domain. 
Third, in the previous work \cite{RDSM:2018,labarbera2020crowdsourcing,SIGIR:2020} the statements judged by the crowd were not recent. This means that evidence on statement truthfulness was often available out there (on the Web), and although the experimental design prevented to easily find that evidence, it cannot be excluded that the workers did find it, or perhaps they were familiar with the particular statement because, for instance, it had been discussed in the press. By focusing on \covid related statements we instead naturally target \emph{recent} statements: 
in some cases the evidence might be still out there, but  this will happen more rarely. 
Fourth, an almost ideal tool to address misinformation would be a crowd able to assess truthfulness in real time, immediately after the statement becomes public: although we are not there yet, and there is a long way to go, we find that targeting recent statements is a step forward in the right direction. 
Fifth, our experimental design differs in some aspects from that used in previous work, and allows us to address novel research questions. 

\section{Background}\label{sec:bg}

\subsection{COVID-19 Infodemic}

The number of initiatives to apply Information Access---and, in general, Artificial Intelligence---techniques to combat the \covid infodemic has been rapidly increasing (see \citet[p.~16]{bullock2020mapping} for a survey).
There is significant effort on analyzing \covid information on social media, and linking to data from external fact-checking organizations to quantify the spread of misinformation 
\cite{gallotti2020assessing,cinelli2020covid19,yang2020prevalence}.
\citet{mejova2020advertisers} analyzed Facebook advertisements related to \covid, and found that around 5\% of them contain errors or misinformation.
Crowdsourcing methodologies have also been used to collect and analyze data from patients with cancer who are affected by the \covid pandemic \cite{desai2020crowdsourcing}. To the best of our knowledge, there is no work  addressing the \covid infodemic using crowdsourcing. 

\subsection{Crowdsourcing Truthfulness}
Recent work has focused on the
automatic classification of truthfulness or  fact checking \cite{Popat_2019, mihaylova2019semeval, atanasova2019automatic, clef2018checkthat, elsayed2019overview,kim2019homogeneity}.
\citet{zubiaga2014tweet} investigated, using crowdsourcing, the reliability of tweets in the setting of disaster management.
CLEF developed a Fact-Checking Lab \cite{clef2018checkthat,elsayed2019overview} to address the issue of ranking sentences according to some fact-checking property.

There is recent work that studies how to collect truthfulness judgments by means of crowdsourcing using fine grained scales \cite{RDSM:2018,labarbera2020crowdsourcing, SIGIR:2020}.
Samples of statements from the PolitiFact dataset---originally published by \citet{politifact}---have been used to analyze the agreement of workers with labels provided by experts in the original dataset. Workers are asked to provide the truthfulness of the selected statements, by means of different fine grained scales. \citet{RDSM:2018} compared two fine grained scales: one in the $[0,100]$ range and one in the $(0,+\infty)$ range, on the basis of Magnitude Estimation \cite{moskowitz1977magnitude}. They found that both scales allow to collect reliable truthfulness judgments that are in agreement with the ground truth. Furthermore, they show that the scale with one hundred levels leads to slightly higher agreement levels with the expert judgments.  On a larger sample of \politifact statements, \citet{labarbera2020crowdsourcing} asked workers to use the original scale used by the \politifact experts and the scale in the $[0,100]$ range. They found that aggregated judgments (computed using the mean function for both scales) have a high level of agreement with expert judgments. Recent work by \citet{SIGIR:2020} found similar results in terms of external agreement and its improvement when aggregating crowdsourced judgments, using statements from two different fact-checkers: \politifact and RMIT ABC Fact Check (\abc).
Previous work has also looked at \emph{internal agreement}, i.e., agreement among workers \cite{RDSM:2018, SIGIR:2020}. \citet{SIGIR:2020} found that scales have  low levels of agreement when compared with each other: correlation values for aggregated judgments on the different scales are around $\rho=0.55-0.6$ for \mbox{\politifact} and $\rho=0.35-0.5$ for \abc, and $\tau=0.4$ for \mbox{\politifact} and $\tau=0.3$ for \mbox{\abc}.  This indicates that the same statements tend to be evaluated differently in different scales. 

There is evidence of differences on the way workers provide judgments, influenced by the sources they examine, as well as the impact of worker bias.
In terms of sources, \citet{labarbera2020crowdsourcing} found that that the vast majority of workers (around 73\% for both scales) use indeed the \politifact website to provide judgments. Differently from \citet{labarbera2020crowdsourcing}, \citet{SIGIR:2020} used a custom search engine in order to filter out \politifact and \abc from the list of results. Results show that, for all the scales, Wikipedia and news websites are the most popular sources of evidence used by the workers.
In terms of worker bias, \citet{labarbera2020crowdsourcing} and \citet{SIGIR:2020} found that worker political background has an impact on how workers provide the truthfulness scores. More in detail, they found that workers are more tolerant and moderate when judging statements from their very own political party.

\section{Aims and Research Questions}\label{sec:rq}

When compared to previous work, in this paper we aim to focus on several novel aspects.
With respect to \citet{RDSM:2018,labarbera2020crowdsourcing,SIGIR:2020}, we focus on claims about \covid, which are recent and interesting for the research community, and are arguably on a more relevant/sensitive topic to the workers. We investigate whether the health domain makes a difference in the ability of crowd workers to identify and correctly classify (mis)information, and if the very recent nature of \covid related statements has an impact as well. 
We focus on a single truthfulness scale, given the 
evidence that the scale used does not make a significant difference \cite{RDSM:2018,labarbera2020crowdsourcing,SIGIR:2020}. 
Another important difference is that we ask workers to provide a textual justification for their decision: we analyze them to better understand the process followed by workers to verify information, and we investigate if they can be exploited to derive useful information.
Finally, we also exploit and analyze worker behavior.

We investigate the following specific Research Questions:
\begin{enumerate}[label=RQ\arabic*]
    
    \item \label{i:RQ1} Are the crowd workers able to detect and objectively categorize online (mis)information related to the medical domain and more specifically to \covid? Which are the relationship and agreement between the crowd and the expert labels?
    
    \item \label{i:RQ2} Can the crowdsourced and/or the expert judgments be transformed or aggregated in a way that it improves the ability of workers to detect and objectively categorize online (mis)information?
    
    \item \label{i:RQ3} Which is the effect of workers' political bias in objectively identifying online misinformation? And the effect of workers' background and Cognitive Reflection Test (CRT) performances?
    
    \item \label{i:RQ4} 
    Which are the signals provided by the workers while performing the task that can be recorded? To what extent are these signals related to workers' accuracy?
    Can these signals be exploited to improve accuracy and, for instance, aggregate the labels in a more effective way?

    \item \label{i:RQ5} Which sources of information does the crowd consider when identifying online misinformation? Are some sources more useful? Do some sources lead to more accurate and reliable assessments by the workers?
    
\end{enumerate}

\section{Methods}\label{sec:methods}
In this section we  present the dataset used to carry out our experiments (Section~\ref{sec:dataset}), and the crowdsourcing task design (Section~\ref{sec:crowd_setup}).
Overall, we considered one dataset annotated by experts, one crowdsourced dataset, 
one judgment scale (the same for the expert and the crowd judgments), and a total of 60 statements.

\subsection{Dataset}\label{sec:dataset}
We considered as primary source of information the \politifact dataset \cite{politifact} that was built as a ``benchmark dataset for fake news detection'' \cite{politifact} and contains over 12k statements produced by public appearances of US politicians. The statements of the datasets are labeled by expert judges on a six-level scale of truthfulness (from now on referred to as \expertsix): \politifactpantsfire, \politifactfalse, \politifactmostlyfalse, \politifacthalftrue, \politifactmostlytrue, and \politifacttrue. 
Recently, the \politifact website (the source from where the statements of the \politifact dataset are taken) created a specific section related to the \covid pandemic.\footnote{\url{https://www.politifact.com/coronavirus/}}
For this work, we selected 10 statements for each of the six \politifact categories, belonging to such a \covid section and with dates ranging from February 2020 to early April 2020. 
Table~\ref{tab:statements} contains some examples of the statements we used. 

\begin{table}[tbp]
\centering
\caption{Examples of \covid fact-checked statements.}
\label{tab:statements}
\begin{adjustbox}{max width=0.47\textwidth}
\begin{tabular}{>{\raggedright}p{5.7cm}p{1cm}p{0.5cm}p{1.9cm}}
\toprule
  \textbf{Statement} & \textbf{Source} & \textbf{Year} & \textbf{Label}\\
 \midrule
  ``We inherited a broken test for COVID-19.'' & Donald Trump & 2020 & \mbox{\politifactpantsfire} \\
 \hline
  ``Church services cannot resume until we are all vaccinated, says Bill Gates.'' & Bloggers & 2020 & \mbox{\politifactfalse}  \\ 
 \hline
  ``Says a 5G law passed while everyone was distracted with the coronavirus pandemic and lists 20 symptoms associated with 5G exposure.'' & Facebook Post & 2020 & \mbox{\politifactmostlyfalse} \\
 \hline
 ``Says a California surfer was alone, in the ocean, when he was arrested for violating the state’s stay-at-home order.'' & Facebook Post & 2020 & \mbox{\politifactmostlytrue}\\ 
 \hline
  ``Photo shows a crowded New York City subway train during stay-at-home order.'' & Viral Image & 2020 & \mbox{\politifacttrue} \\ 
\bottomrule
\end{tabular}
\end{adjustbox}
\end{table}

\subsection{Crowdsourcing Experimental Setup}\label{sec:crowd_setup}
To collect our judgments we used the crowdsourcing platform Amazon Mechanical Turk (MTurk).
Each worker, upon accepting our HIT, is redirected to an external server to complete the task; we set the payment to \$1.5 for a set of 8 statements\footnote{Before deploying the task on MTurk, we investigated the average time spent to complete the task, and we related it to the minimum US hourly wage.}.
The task itself is as follows: first, a (mandatory) questionnaire is shown to the worker, to collect his/her background information such as age and political views. Then, the worker needs to provide answers to three Cognitive Reflection Test (CRT) questions, which are used to measure the personal tendency to answer with an incorrect ``gut'' response or engage in further thinking to find the correct answer \cite{Frederick2005}.\footnote{We used the same CRT settings as \citet{SIGIR:2020}.}
After the questionnaire and CRT phase, the worker is asked to asses the truthfulness of 8 statements: 6 from the dataset described in \ref{sec:dataset} (one for each of the six considered \politifact categories) and 2 special statements called \emph{Gold Questions}, one clearly true and the other clearly false,  manually written by the authors of this paper and used as quality checks.
We used a randomization process when building the HITs to avoid all the possible source of bias, both  within each HIT and considering the overall task. 

To assess the truthfulness of each statement, the worker is shown: the \emph{Statement}, the \emph{Speaker/Source}, and the \emph{Year} in which the statement was made.
We asked the worker to provide the following information: 
the \emph{truthfulness value} for the statement using the six-level scale adopted by \politifact, from now on referred to as \six (presented to the worker using a radio button containing the label description for each category as reported in the original \politifact website),
a \emph{URL} that s/he used as a source of information for the fact checking, and 
a textual \emph{motivation} for her/his response (which can not include the URL, and should contain at least 15 words).
In order to prevent the user from using \politifact as primary source of evidence, we implemented a custom search engine, which is based on the Bing Web Search APIs\footnote{https://azure.microsoft.com/services/cognitive-services/bing-web-search-api/} and filters out \politifact from the returned search results.

We logged the user behavior using a custom logger \cite{han2019all,8873609}, and we implemented in the task the following quality checks:
(i) the judgments assigned to the gold questions have to be coherent (i.e., the judgment of the clearly false question should be lower than the one assigned to true question); and (ii) the cumulative time spent to perform each judgment should be of at least 10 seconds.
Note that the CRT (and the questionnaire) answers were not used for quality check, although the workers were not aware of that.

Overall, we used 60 statements in total (10 for each  \politifact category), and each statement has been evaluated by 10 distinct workers. Thus, we deployed 100 MTurk HITs and we collected 800 judgments in total.
The crowd task was launched on  May 1st, 2020 and it completed on May 4th, 2020.
All the data used to carry out our experiments can be downloaded at \url{https://github.com/KevinRoitero/crowdsourcingTruthfulness}.

\section{Results and Analysis}\label{sec:analysis}
We first
report some descriptive statistics 
 about the population of workers and the data collected in our experiment
 (Section \ref{sec:descriptive}).
Then, we address crowd accuracy (i.e., \ref{i:RQ1}) in Section \ref{sec:accuracy},
transformation of truthfulness scales (\ref{i:RQ2}) in Section \ref{sec:transforming_scales},
worker background and bias (\ref{i:RQ3}) in Section \ref{sec:worker_background_and_bias}, 
worker behavior (\ref{i:RQ4}) in Section \ref{sec:worker_behavior}; finally, we study 
  information sources  (\ref{i:RQ5}) in Section \ref{sources_of_information}.

\subsection{Descriptive Statistics}\label{sec:descriptive}

\subsubsection{Worker Background, Abandonment, and Bias}\label{sec:worker_background}
\leavevmode
\phantom{a}

\myparagraph{Questionnaire}
Overall, \num{1113} workers resident in the United States
participated in our experiment.\footnote{Workers provide proof that they are based in US and have the eligibility to work.}
In each HIT, workers were first asked to complete a demographics questionnaire with questions about their gender, age, education and political views.
By analyzing the answers to the questionnaire 
we derived the following demographic statistics. 
The majority of workers are in the 26--35 age range (44\%), followed by 36--50 (25\%), and 19--25 (18\%).
The majority of workers are well educated: 47\% of them have a four year college degree or a bachelor degree, 21\% have a college degree, and 17\% have a postgraduate or professional degree. Only about 15\% of workers have a high school degree or less. 
Concerning political views, we had 28\% of workers that identified themselves as liberals, 28\% as moderate, 24\% as conservative, 11\% as very conservative and 9\% as very liberal. Moreover, 44\% of workers identified themselves as being Democrat, 31\% as being Republican, and 22\% as being Independent. Finally, 46\% of workers agree on building a wall on the southern US border, and 42\% of them disagree.
Overall we can say that 
our sample is  well balanced. 

\myparagraph{CRT Test}\label{sec:crt_test}
Analyzing the CRT scores, we found that:
31\% of workers did not provide any correct answer, 
34\% answered correctly to 1 test question, 
18\% answered correctly to 2 test questions, and only
17\% answered correctly to all 3 test questions.

\myparagraph{Abandonment}\label{sec:abandonment}
When considering the abandonment ratio (measured according to the definition provided by \citet{han2019all,8873609}), we found that \num{100} of the workers (about 9\%) successfully completed the task, \num{991} (about 87\%) abandoned (i.e., voluntarily terminated the task before completing it), and \num{45} (about 4\%) failed (i.e., terminated the task due to failing the quality checks too many times).
Most of the abandonment 
(80\% of the \num{1091} workers, 85\% of the \num{991} workers that abandoned) happened before judging the first statement (i.e., before really starting the task);
about 7.52\% of the \num{1091} workers (8\% of the \num{991} of the workers that abandoned) abandoned after the last statement (most likely once failed the quality check).

\subsubsection{Crowdsourced Scores}\label{sec:crowd_score_distribution}

\begin{figure}[tbp]
  \centering
  \begin{tabular}{@{}c@{}c@{}c@{}}
    \includegraphics[width=.35\linewidth]{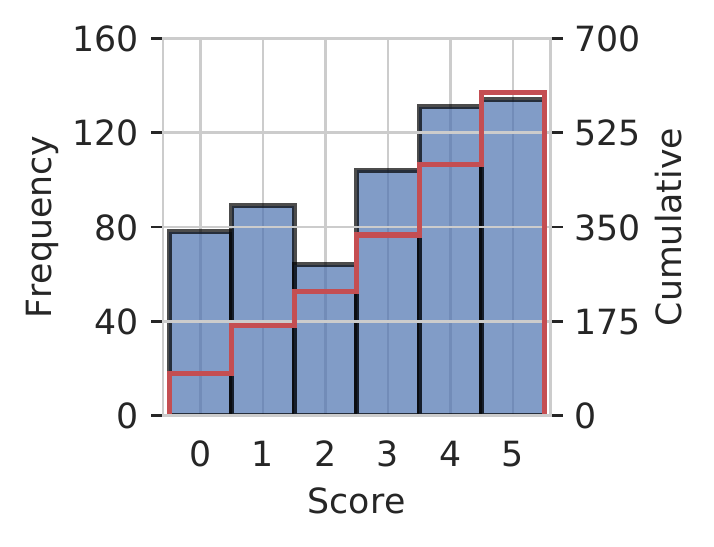}&
    \includegraphics[width=.3\linewidth]{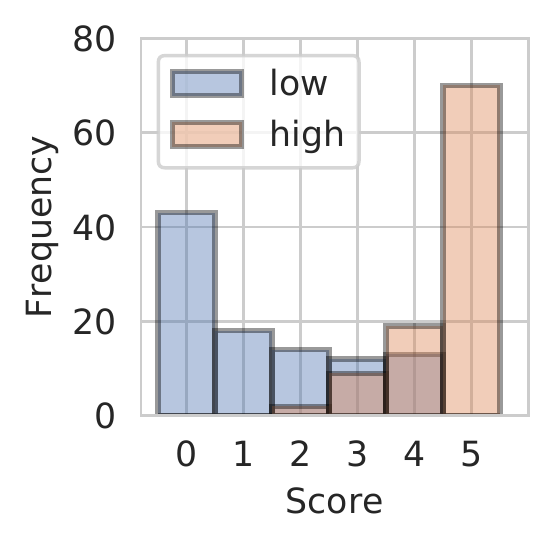}&
    \includegraphics[width=.35\linewidth]{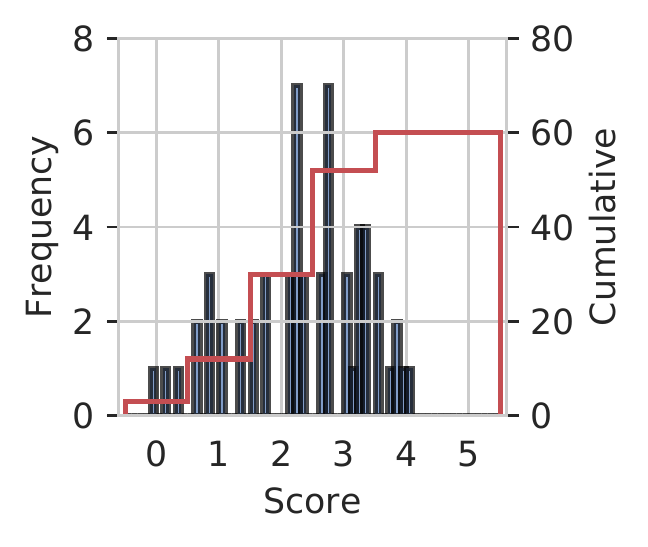}\\
  \end{tabular}
\caption{%
     Distribution (in blue) and the cumulative distribution (in red) of the individual (left), gold (middle), and aggregated with mean (right). 
}
  \label{fig:scores_distributions}
\end{figure}

Figure~\ref{fig:scores_distributions}  
shows the distribution (in blue) and the cumulative distribution (in red) of the individual (left), gold (middle), and aggregated with mean (right) scores provided by the workers for the considered \politifact statements.

If we focus on the distribution of the individual scores (left plot), we can see that the distribution is quite well balanced, just lightly skewed towards higher truthfulness values, represented in the rightmost part of the plot. This behavior is also remarked when focusing on the red line representing the cumulative distribution, which displays almost evenly spaced steps.
This is a first indication that suggests that crowd judgments are overall of a decent quality; in fact, our empirical distribution is not distant from the ideal one: since we considered 10 statements for each \politifact category, the perfect distribution would have been the uniform distribution.

Turning to the distribution of the gold scores (i.e., the two special statements used for quality check, shown in the middle plot), we see that the large majority of workers (i.e., 70\% for \high and 43\% for \low) used the extreme values of the scale (i.e., \politifactpantsfire and \politifacttrue); furthermore, we see that overall the \high gold question 
has been judged correctly more times than the \low gold question, 
suggesting the probably the workers found the former easier to judge than the latter.

We now turn to analyze the distribution of the scores when aggregated using the mean function (shown in the right plot). The distribution for the aggregated scores becomes roughly bell-shaped, and slightly skewed towards high truthfulness values---this behavior is consistent with the findings of \citet{SIGIR:2020}.
In the following we discuss both the external (i.e., between workers and experts) and internal (i.e.,  among workers) agreement of our dataset.

\begin{figure*}[tbp]
  \centering
  \begin{tabular}{@{}c@{}c@{}c@{}c@{}}
      \includegraphics[width=.25\linewidth]{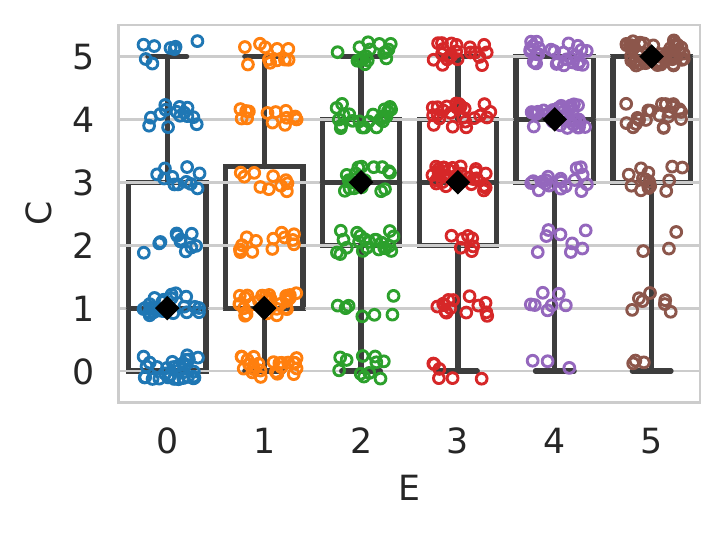}&
    \includegraphics[width=.25\linewidth]{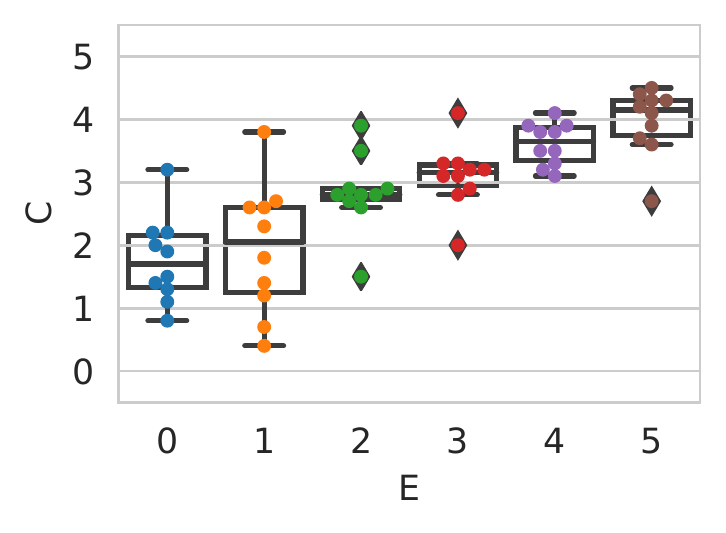}&
    \includegraphics[width=.25\linewidth]{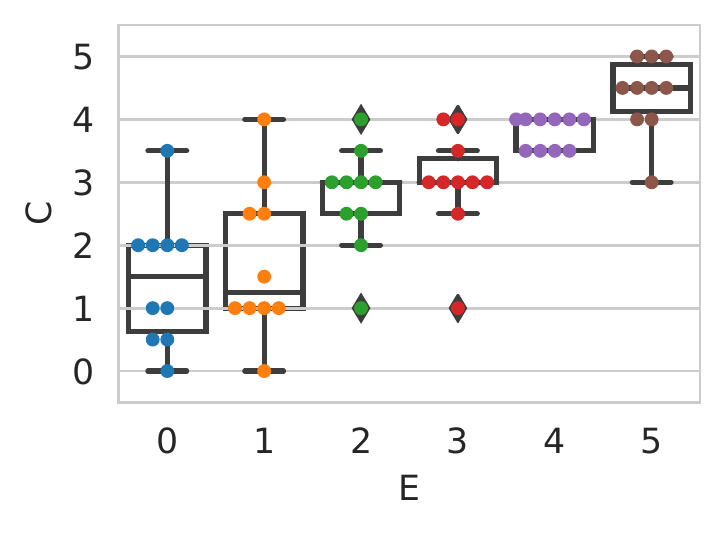}&
    \includegraphics[width=.25\linewidth]{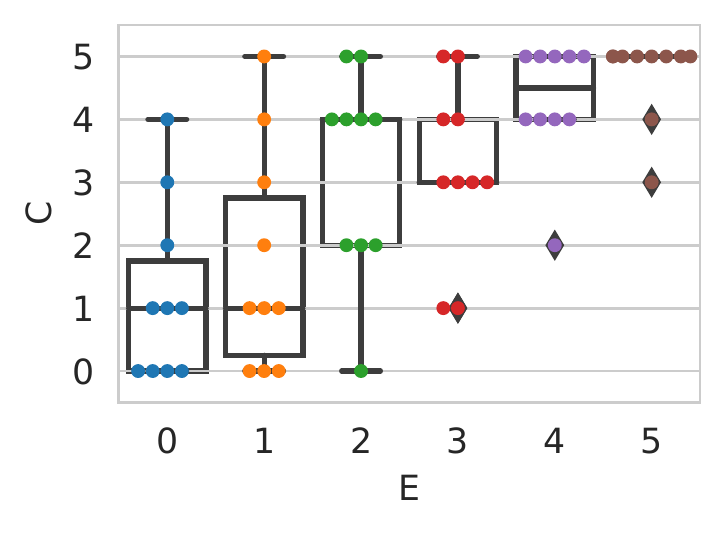}
  \end{tabular}\vspace*{-2mm}
\caption{%
The agreement between the \politifact experts (x-axis) and the crowd judgments (y-axis).
From left to right: 
 \six individual judgments; 
 \six aggregated with mean; 
 \six aggregated with median;
 \six aggregated with majority vote. 
}
  \label{fig:agreement_ground_truth}
\end{figure*}

\subsection{\ref{i:RQ1}: Crowd Accuracy}\label{sec:accuracy}

\subsubsection{External Agreement}
To answer \ref{i:RQ1}, we start by analyzing the so called external agreement, i.e., the agreement between the crowd collected labels and the experts ground truth.
Figure~\ref{fig:agreement_ground_truth} shows the agreement between the \politifact experts (x-axis) and the crowd judgments (y-axis). In the first plot, each point is a judgment by a worker on a statement, i.e., there is no aggregation of the workers working on the same statement. 
In the next plots all workers redundantly working on the same statement are aggregated using the mean (second plot), median (third plot), and majority vote (right-most plot). 
If we focus on the first plot (i.e., the one with no aggregation function applied), we can see that, overall, the individual judgments are in agreement with the expert labels, as shown by the  median values of the boxplots, which are increasing as the ground truth truthfulness level increases.
Concerning the aggregated values, it is the case that for all the aggregation functions the \politifactpantsfire and \politifactfalse categories are perceived in a very similar way by the workers; this behavior was already shown in previous work \cite{SIGIR:2020, labarbera2020crowdsourcing}, and suggests that indeed workers have clear difficulties in distinguishing between the two categories; this is even more evident considering that the interface presented to the workers contained a textual description of the categories' meaning in every page of the task. 

If we look at the plots as a whole, we see that within each plot the median values of the boxplots are increasing when going from \politifactpantsfire to \politifacttrue (i.e., going from left to right of the x-axis of each chart). This indicates that the workers are overall in agreement with the \politifact ground truth, thus indicating that workers are indeed capable of recognizing and correctly classifying misinformation statements related to the \covid pandemic. This is a very important and not obvious result: in fact, the crowd (i.e., the workers) is the primary source and cause of the spread of disinformation and misinformation statements across social media platforms \cite{chen2015students}.
By looking at the plots, and in particular focusing on the median values of the boxplots, it appears evident that the mean (second plot) is the aggregation function which leads to higher agreement levels, followed  by the median (third plot) and the majority vote (fourth plot). Again, this behavior has already been noticed \cite{SIGIR:2020, labarbera2020crowdsourcing, Roitero:2018:FRS:3209978.3210052}, and all the cited works used the mean as primary aggregation function.

To validate the external agreement, we measured the statistical significance between the aggregated rating for all the six \politifact categories; we considered both the  Mann-Whitney rank test and the t-test, applying Bonferroni correction to account for multiple comparisons. Results are as follows:
when considering adjacent categories (e.g., \politifactpantsfire and \politifactfalse), the difference between categories are never significant, for both tests and for all the three aggregation functions. 
When considering categories of distance 2 (e.g., \politifactpantsfire and \politifactmostlyfalse), the differences are never significant, apart from the median aggregation function, where there is statistical significance to the $p<.05$ level in $2/4$ cases for both Mann-Whitney and t-test. 
When considering categories of distance 3, the differences are significant,
for the mean, in $3/3$ cases for the Mann-Whitney and $3/3$ cases for the t-test,
for the median,  in $2/3$ cases for the Mann-Whitney and $3/3$ cases for the t-test,
for the majority vote, in $0/3$ cases for the Mann-Whitney and $1/3$ cases for the t-test.
When considering categories of distance 4 and 5, the differences are always significant to the $p>0.01$ level for all the aggregation functions and for all the tests, apart from the case of the majority vote function and the Mann-Whitney test, where the significance is at the $p>.05$ level. 
In the following we use the mean as being the most commonly used approach for this type of data \cite{SIGIR:2020}.

\begin{figure}[tbp]
  \centering
  \begin{tabular}{@{}c@{}}
   \includegraphics[width=.79\linewidth]{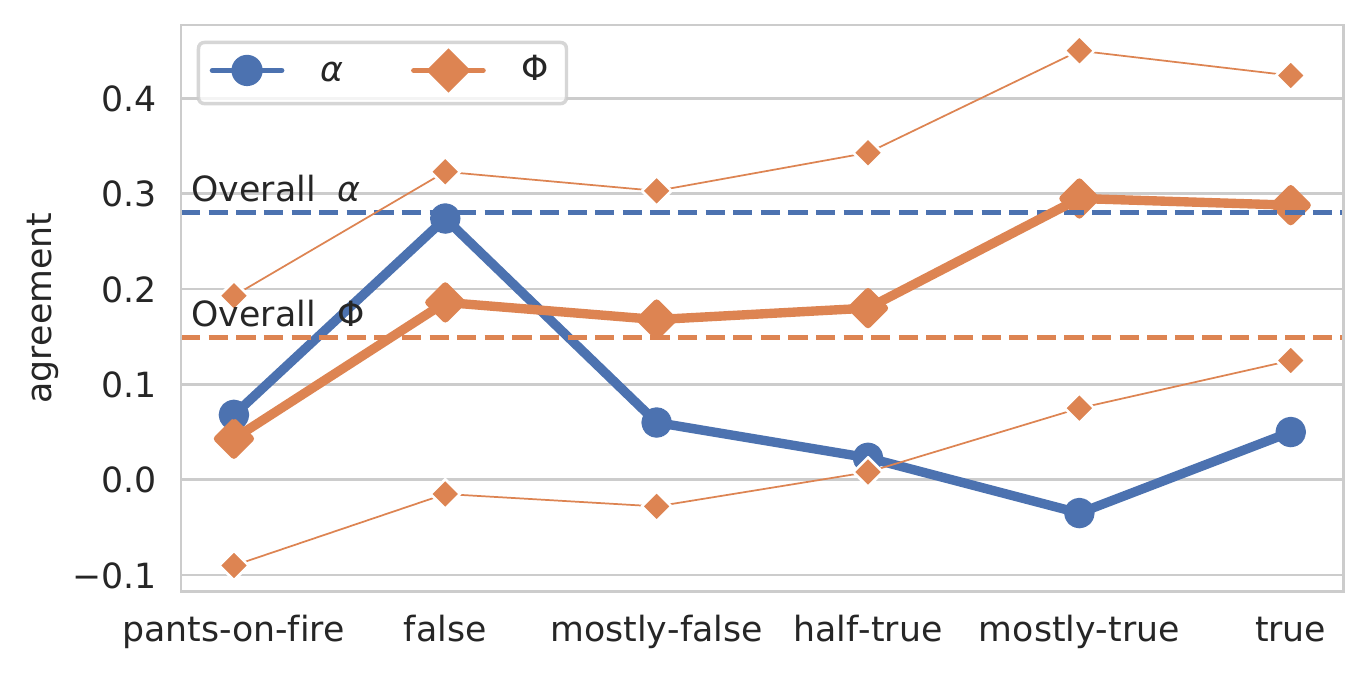}\\
    \includegraphics[width=.79\linewidth]{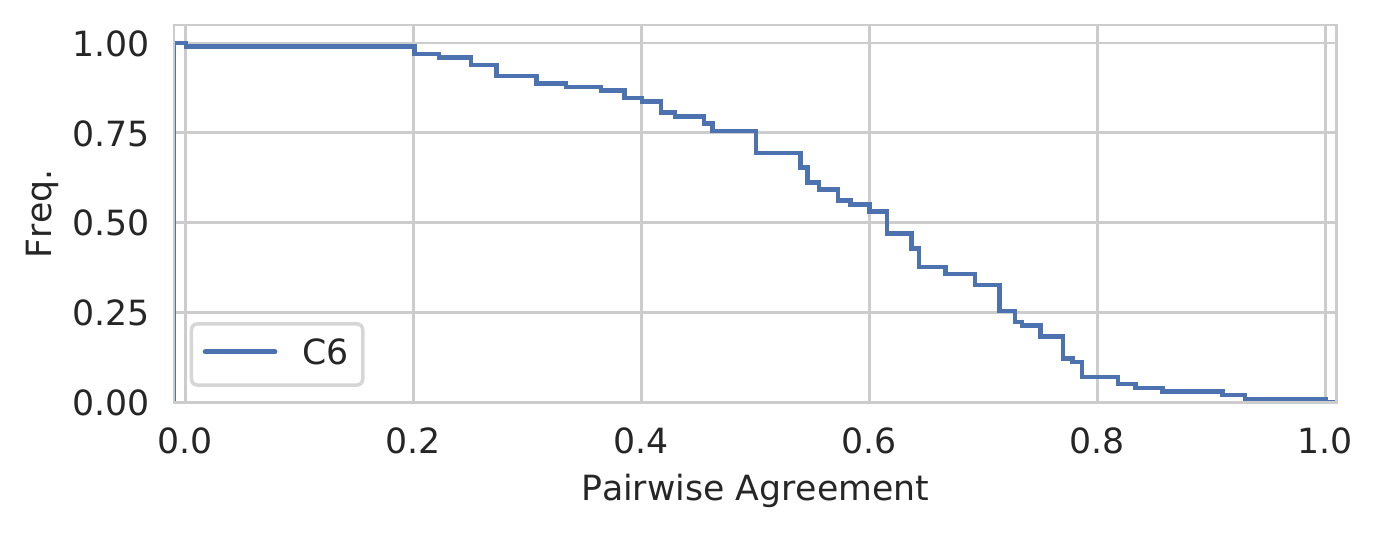}\\
  \end{tabular}
\caption{%
    Workers agreement: $\alpha$ \cite{krippendorff2011computing} and $\Phi$ \cite{checco2017let} (top plot); pairwise unit agreement (bottom plot). 
}
  \label{fig:agreement}
\end{figure}

\subsubsection{Internal Agreement}
 Another standard way to address~\ref{i:RQ1} and to analyze the quality of the work by the crowd is to compute the so called internal agreement (i.e., the agreement among the workers). 
Figure~\ref{fig:agreement} shows in the first plot the agreement measured with $\alpha$ \cite{krippendorff2011computing} and $\Phi$ \cite{checco2017let}, two popular measures often used to compute workers' agreement in crowdsourcing tasks \cite{RDSM:2018, SIGIR:2020, 10.1145/3121050.3121060, Roitero:2018:FRS:3209978.3210052}. The x-axis details the \politifact categories, while the y-axis the level of agreement measured; while $\alpha$ is a punctual measure, $\Phi$ allows to compute confidence intervals for the  agreement measure; the plot shows the upper 97\% and lower 3\% confidence intervals as thinner lines.
As we can see from the plot, the agreement levels measured with the two scales is very similar for the \politifactpantsfire, \politifactfalse, \politifactmostlyfalse, and \politifacthalftrue categories: note that the $\alpha$ measure always falls in the $\Phi$ confidence interval, and the 
little oscillations in the agreement value are not always indication of a real change in the agreement level, especially when considering $\alpha$ \cite{checco2017let}. Having said so, it appears that for all the two metrics the overall agreement falls in the $[0.15, 0.3]$ range, and the agreement level is similar for all the \politifact categories, with the exception of $\Phi$, which shows higher agreement levels for the \politifactmostlytrue and \politifacttrue categories. This confirms the finding, derived from Figure~\ref{fig:agreement_ground_truth},  that workers seem most effective in identifying and categorizing statements with a higher truthfulness level. This remark is also supported by \cite{checco2017let} which shows that $\Phi$ is better in distinguishing agreement levels in crowdsourcing than $\alpha$, which is more indicated as a measure of data reliability in non crowdsourced settings.

Figure~\ref{fig:agreement} also shows in  the second plot a measure of the agreement at the HIT level (i.e., in the set of 8 statements judged by each worker) as detailed in \cite{10.1145/3121050.3121060, SIGIR:2020}. 
More in detail, the plot shows the CCDF (Complementary Cumulative Distribution Function) of the relative frequencies for the agreement of the 100 HITs considered in this experiment. The plot shows that around 20\% of the hits have a pairwise agreement which is very close to 1; this indicates that around 20\% of the workers judged statements almost in the same way as the expert judges. Moreover, we see that 60\% of the workers have a pairwise agreement greater than $0.5$. Again, this result indicates a good overall agreement between crowd and expert judgments, confirming that the crowd is able to correctly identify and classify misinformation related to the \covid pandemic.

\subsection{\ref{i:RQ2}: Transforming Truthfulness Scales} \label{sec:transforming_scales}
Given the positive results presented above, it appears that the answer to \ref{i:RQ1} is overall positive, even if with some exceptions. 
There are many remarks that can be made:
first, there is a clear issue that affects the \politifactpantsfire and \politifactfalse categories, which are very often mis-classified by workers. Moreover, while \politifact used a six-level judgment scale, the usage of a two- (e.g., True/False) and a three-level (e.g., False / In between / True) scale is also common when assessing the truthfulness of statements \cite{labarbera2020crowdsourcing, SIGIR:2020}.
Finally, categories can be merged together to improve accuracy, as done for example by \citet{tchechmedjiev2019claimskg}.
All these considerations lead us to \ref{i:RQ2}, addressed in the following.

\subsubsection{Merging Ground Truth Levels}
For all the above reasons, we performed the following experiment: we group together the six \politifact categories (i.e., \expertsix) into three (referred to as \expertthree) or two (\experttwo) categories, which we refer  respectively with \politifactthreebinszero, \politifactthreebinsone, and \politifactthreebinstwo for the three level scale, and 
\politifacttwoebinszero and \politifacttwobinsone for the two level scale. 

\begin{figure}[tbp]
  \centering
  \begin{tabular}{@{}c@{}c@{}c@{}}
     \includegraphics[width=.33\linewidth]{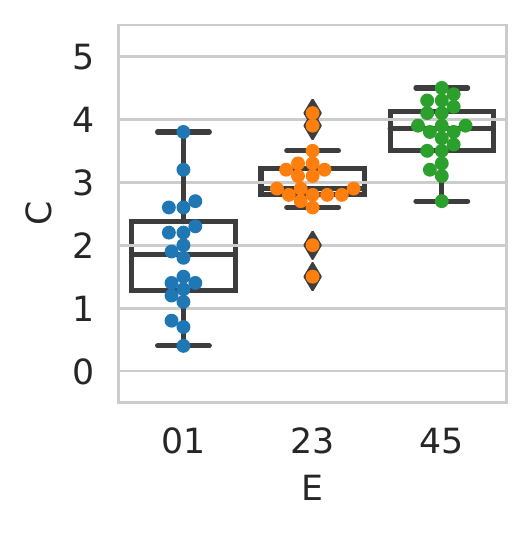}&
    \includegraphics[width=.33\linewidth]{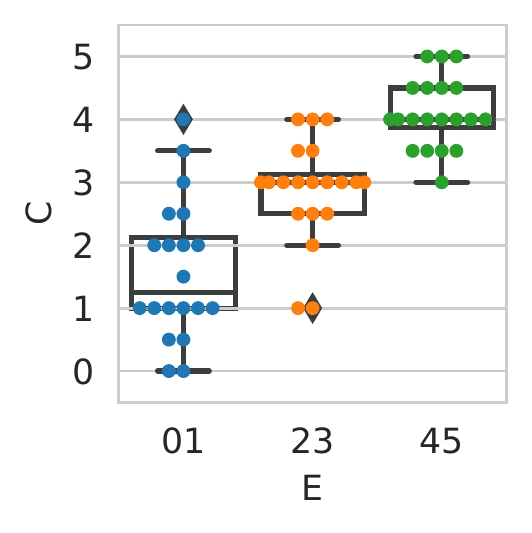}&
    \includegraphics[width=.33\linewidth]{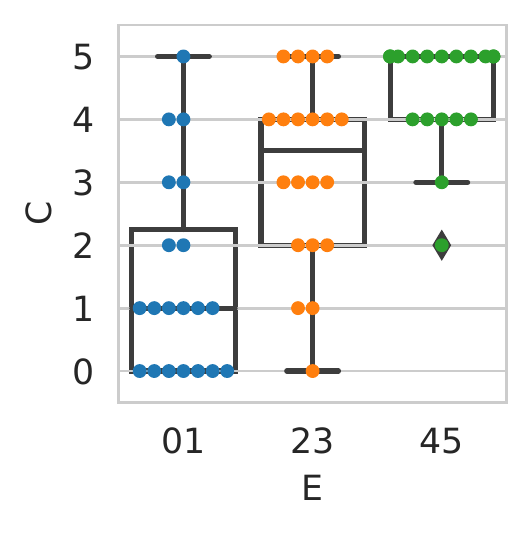}\\
     \includegraphics[width=.33\linewidth]{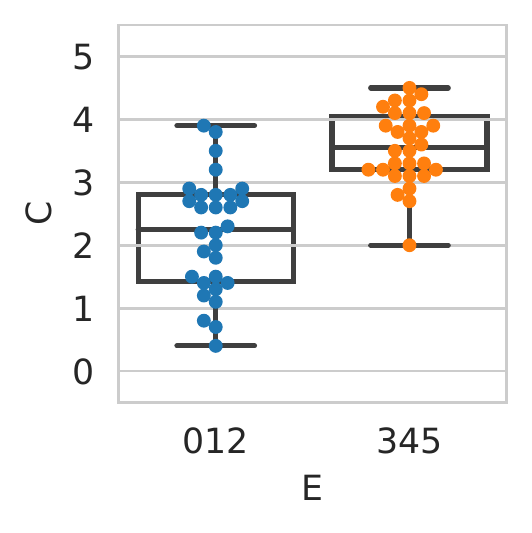}&
    \includegraphics[width=.33\linewidth]{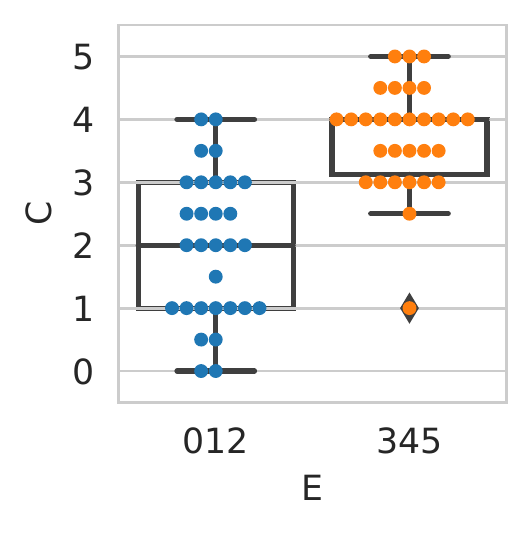}&
    \includegraphics[width=.33\linewidth]{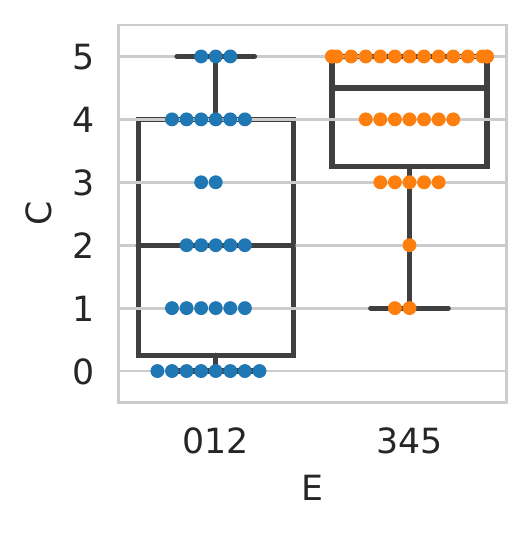}
  \end{tabular}
\caption{%
    The agreement between the \politifact experts and the crowd judgments. From left to right: 
    \six aggregated with mean;
    \six aggregated with median; 
    \six aggregated with majority vote.
    First row: \expertsix to \expertthree; 
    second row: \expertsix to \experttwo. 
    Compare with Figure~\ref{fig:agreement_ground_truth}.
}
  \label{fig:binning}
\end{figure}

Figure~\ref{fig:binning} shows the result of such a process. As we can see from the plots, the agreement between the crowd and the expert judgments can be seen in a more neat way. As for Figure~\ref{fig:agreement_ground_truth}, the median values for all the boxplots is increasing when going towards higher truthfulness values (i.e., going from left to right within each plot); this holds for all the aggregation functions considered, and it is valid for both transformations of the \expertsix scale, into two and three levels.
Also in this case we computed the statistical significance between categories, applying the Bonferroni correction to account for multiple comparisons. Results are as follows.
For the case of three groups, both the categories at distance one and two are always significant to the $p<0.01$ level, for both the Mann-Whitney and the t-test, for all three aggregation functions.
The same behavior holds for the case of two groups, where the categories of distance 1 are always significant to the $p<0.01$ level.

Summarizing, we can now conclude that by merging the ground truth levels we obtained a much stronger signal: the crowd can effectively detect and classify misinformation statements related to the \covid pandemic.

\subsubsection{Merging Crowd Levels}
Having reported the results on merging the ground truth categories we now turn to transform the crowd labels (i.e., \six) into 
three (referred to as \three) 
and 
two (\two) categories.
For the transformation process we rely on the approach detailed by \citet{scale}, that also present a 
complete and exhaustive discussion on the effectiveness of the scale transformation methods.
This approach has many advantages \cite{scale}: we can simulate the effect of having the crowd answers in a more coarse-grained scale (rather than \six), and thus we can simulate new data without running the whole experiment on MTurk again. 
As we did for the ground truth scale, we choose to select as target scales the two- and three- levels scale, driven by the same motivations.
Having selected \six as being the source scale, and having selected the target scales as the three- and two- level ones (\three and \two), we perform the following experiment. We perform all the possible cuts\footnote{\six can be transformed into \three in 10 different ways, and \six can be transformed into \two in 5 different ways.} from \six to \three  and from \six to \two; then, we measure the internal agreement (using $\alpha$ and $\Phi$) both on the source and on the target scale, and we compare those values. In such a way, we are able to identify, among all the possible cuts, the cut which leads to  the highest possible internal agreement. Also in this case, a detailed discussion on the relationships between internal agreement, effectiveness, and all the possible cuts can be found in \citet{scale}.

\begin{figure}[tbp]
  \centering
  \begin{tabular}{@{}c@{}c@{}}
   \includegraphics[width=.48\linewidth]{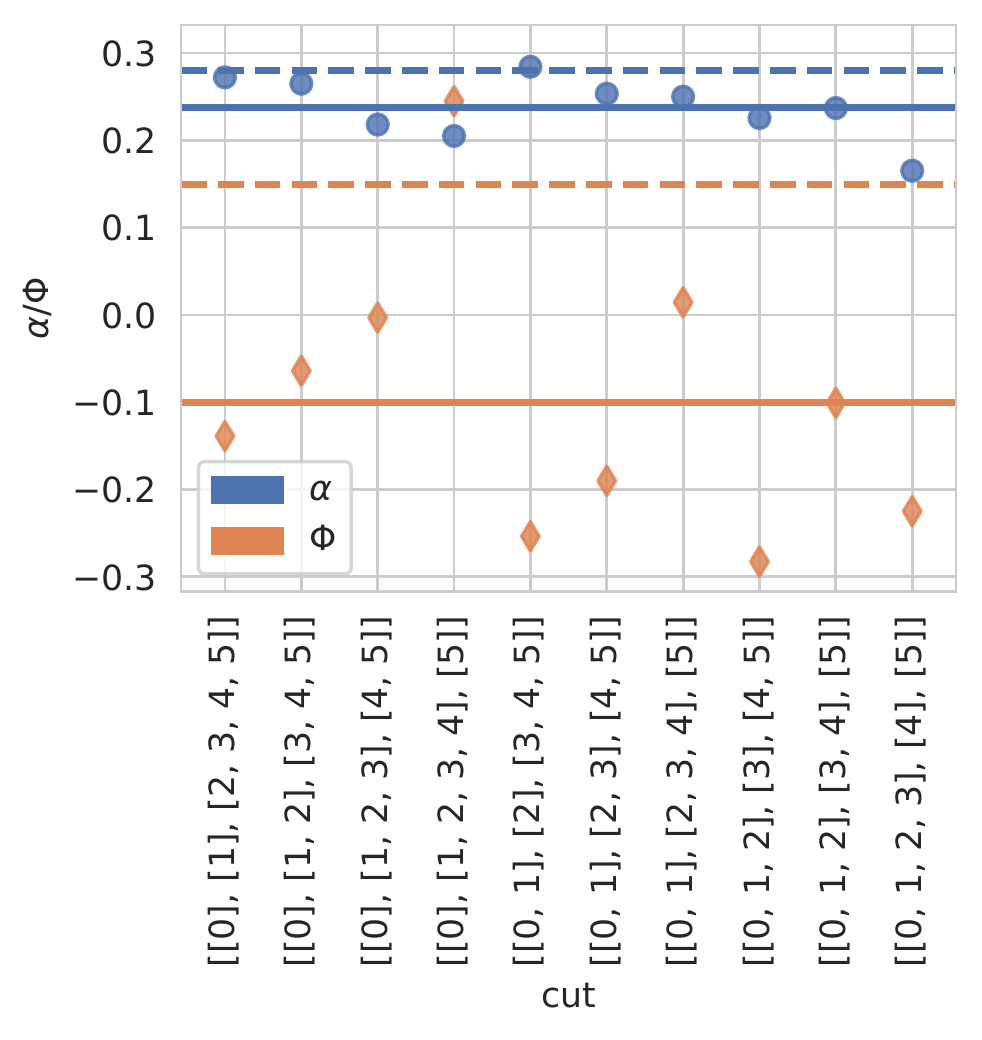}&
    \includegraphics[width=.48\linewidth]{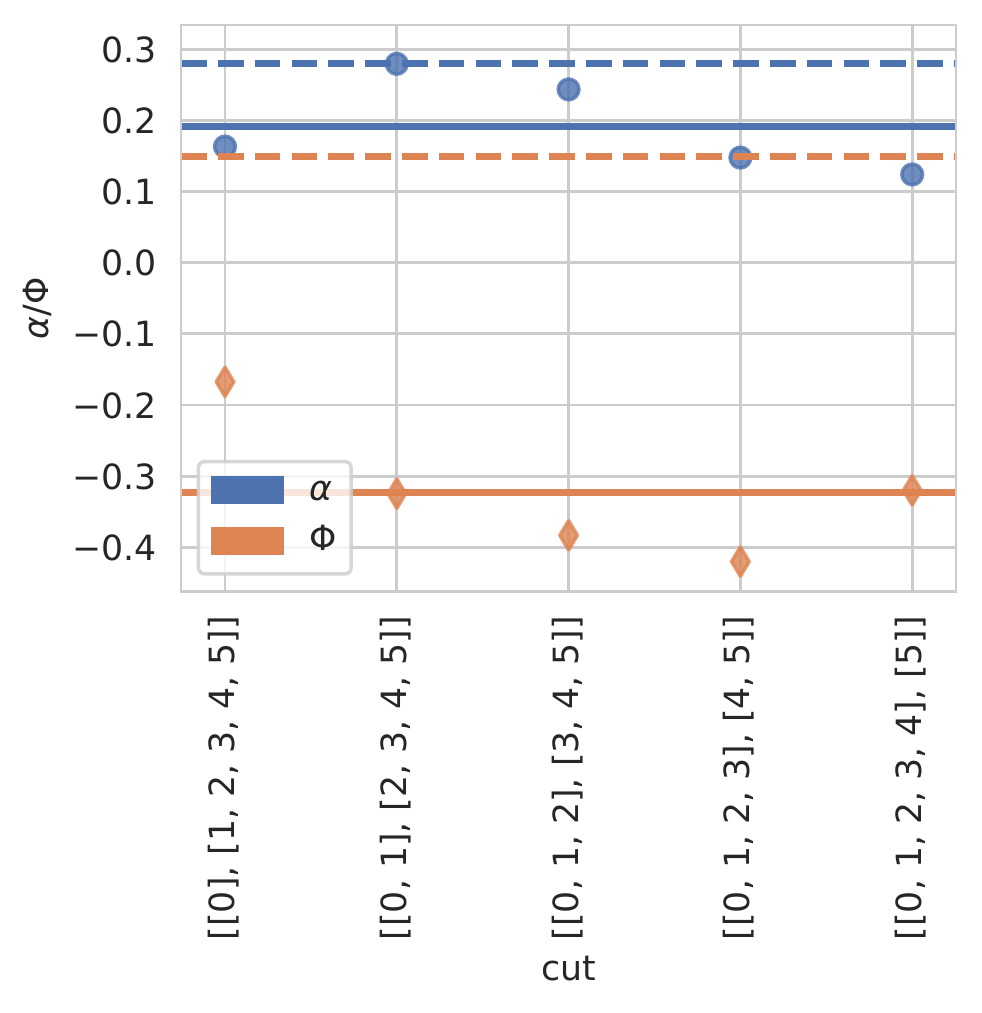}\\
    
  \end{tabular}
\caption{%
     $\alpha$ and $\Phi$ cuts.  
    From left to right: 
    \six cut into three levels (\three),
    \six cut into two levels (\two).
    The cut is detailed in the x-label.
    The dotted line is $\alpha$ / $\Phi$ on \six, 
    the continuous line is the average $\alpha$ / $\Phi$ score measured among all the cuts.
}
  \label{fig:trans_agreement}
\end{figure}

Figure~\ref{fig:trans_agreement} shows the results. The x-axis  shows the cut performed to transform \six into the target scale (\three in the left-most plot, \two in the right-most plot), while the y-axis shows the internal agreement score by means of either $\alpha$ or $\Phi$. As we can see by inspecting the left-most plot, (i.e., \six to \three) we can see that there is, both for  $\alpha$ and $\Phi$, a single cut which leads to higher agreement levels with the original \six scale. On the contrary, if we focus on the rightmost plot (i.e., \six to \two), we can see that there is a single cut for $\alpha$ which leads to similar agreement levels as in the original \six scale, and there are no cuts with such a property when using $\Phi$. 

Having identified the best possible cuts for both transformations and for both agreement metrics, we now measure the external agreement between the crowd and the expert judgments, using the selected cut. 
\begin{figure}[tbp]
  \centering
  \begin{tabular}{@{}c@{}c@{}}
     \includegraphics[width=.49\linewidth]{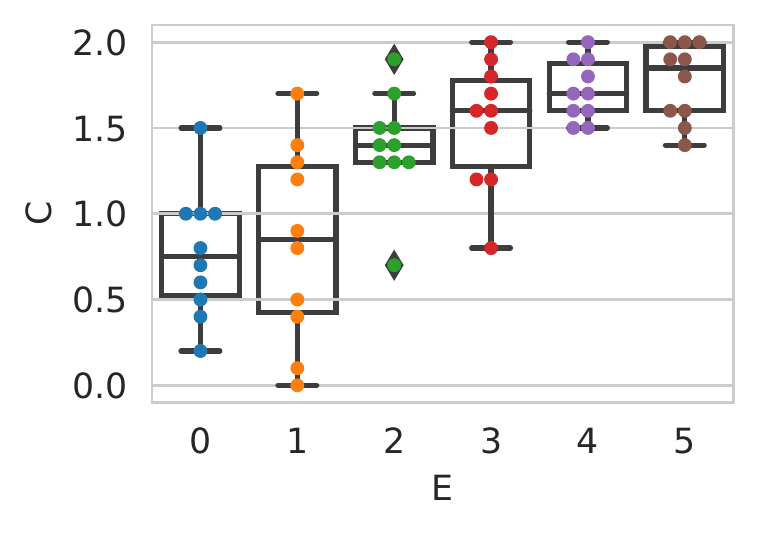}&
    \includegraphics[width=.49\linewidth]{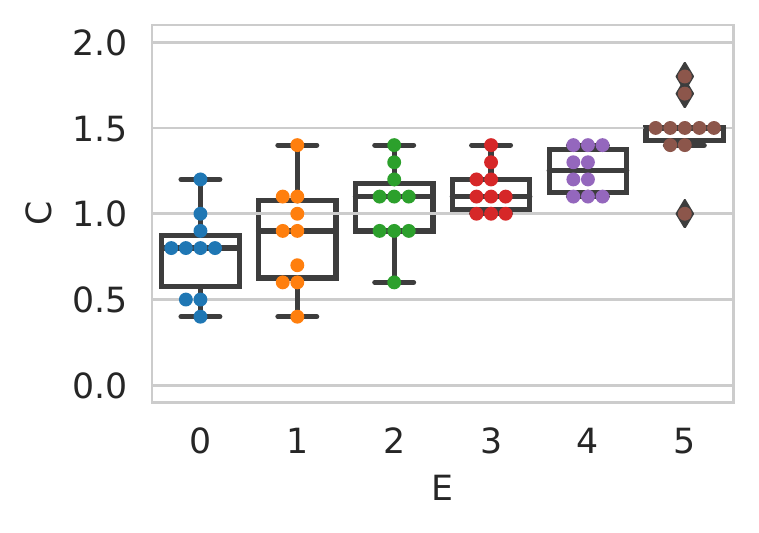}\\
     \includegraphics[width=.49\linewidth]{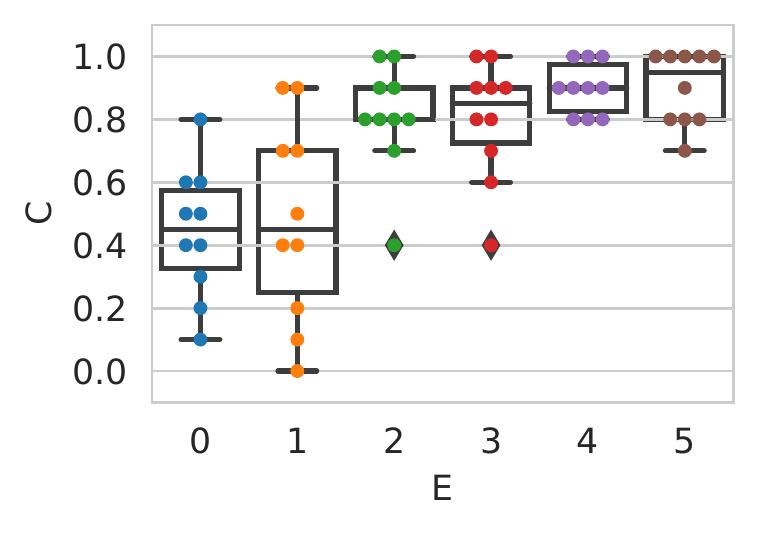}&
    \includegraphics[width=.49\linewidth]{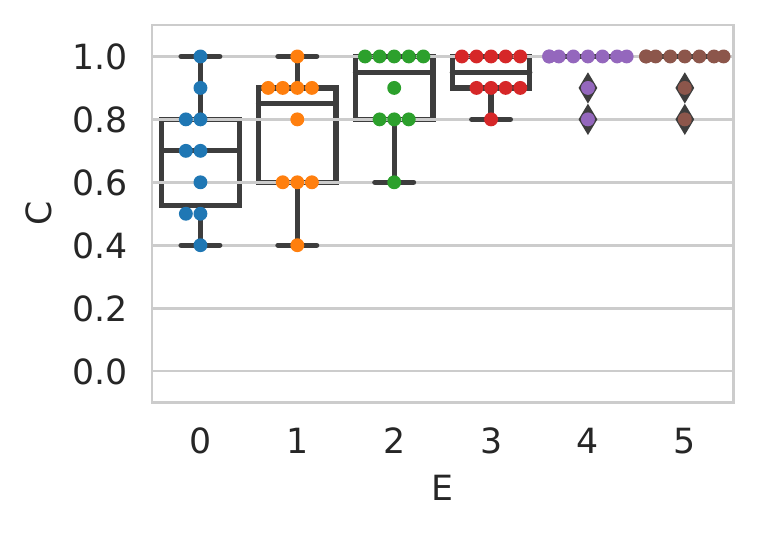}
  \end{tabular}
\caption{%
    Comparison with \expertsix.
    \six to \three (first row)  and to \two (second row), then aggregated with the mean function. 
    Best cut selected according to $\alpha$ (left column) and $\Phi$ (right column) (see Figure~\ref{fig:trans_agreement}). 
    Compare with Figure~\ref{fig:agreement_ground_truth}.
}
  \label{fig:transforming_scales}
\end{figure}
Figure~\ref{fig:transforming_scales} shows such a result when considering the judgments aggregated with the mean function. As we can see from the plots, it is again the case that the median values of the boxplots is always increasing, for all the transformations. Nevertheless, inspecting the plots we can state that the overall external agreement appears to be lower than the one shown in Figure~\ref{fig:agreement_ground_truth}. Moreover, we can also state that also in the case of the transformed scales, the categories \politifactpantsfire and \politifactfalse are still not separable. 
Summarizing, we show that it is feasible to transform the judgments collected on a \six level scale into two new scales,  \three and \two, obtaining judgments with a similar internal agreement as the original ones,  and with a slightly lower external agreement with the expert judgments.

\subsubsection{Merging both Ground Truth and Crowd Levels}
\begin{figure}[tbp]
  \centering
  \begin{tabular}{@{}c@{}c@{}}
     \includegraphics[width=.49\linewidth]{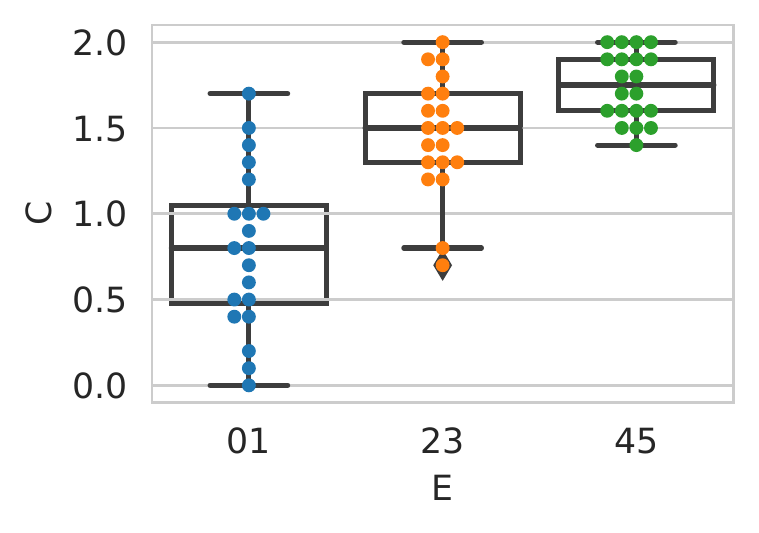}&
    \includegraphics[width=.49\linewidth]{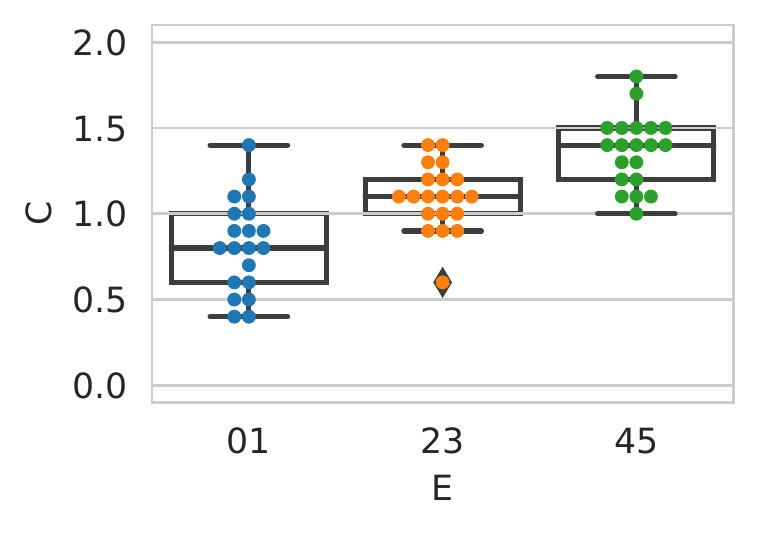}\\
     \includegraphics[width=.49\linewidth]{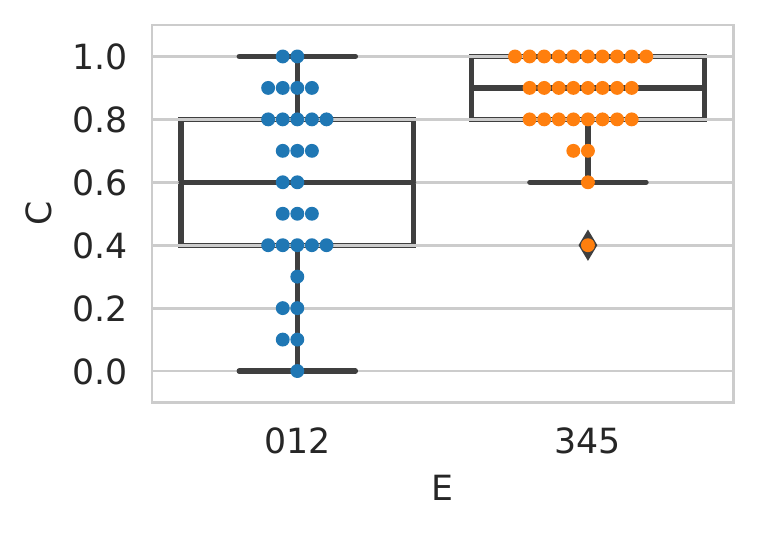}&
    \includegraphics[width=.49\linewidth]{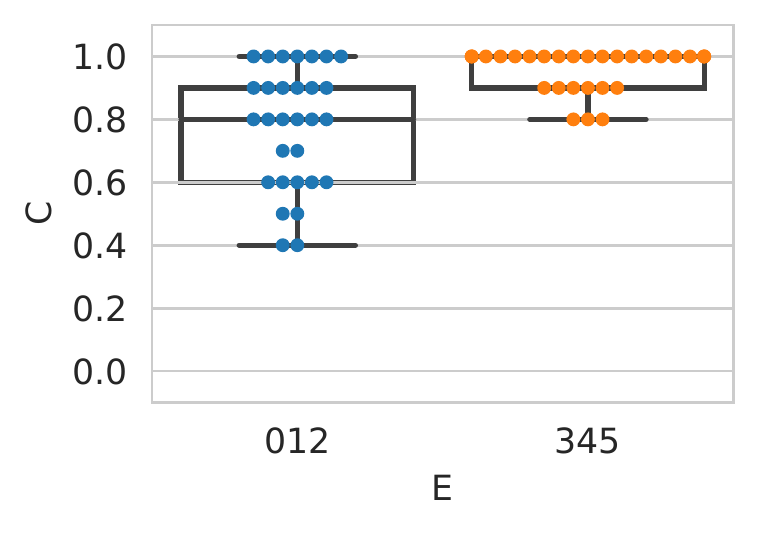}
  \end{tabular}
\caption{%
    \six to \three (first row)  and to \two (second row), then aggregated with the mean function. 
    First row: \expertsix to \expertthree.
    Second row: \expertsix to \experttwo. 
    Best cut selected according to $\alpha$ (left column) and $\Phi$ (right column) (see Figure~\ref{fig:trans_agreement}). Compare with Figures~\ref{fig:agreement_ground_truth}, \ref{fig:binning}, and ~\ref{fig:transforming_scales}. 
}
  \label{fig:transforming_scales_binning}
\end{figure}

It is now natural to combine the two approaches. Figure~\ref{fig:transforming_scales_binning} shows the comparison between 
\six transformed into  \three and \two, and
\expertsix transformed into \expertthree and  \experttwo. 
As we can see form the plots, also in this case the median values of the boxplots are increasing, especially for the \expertthree case (shown in the first row). Furthermore, the external agreement with the ground truth is present, even if for the \experttwo case (shown in the second row) the classes appear to be not separable.
Summarizing, all these results show that it is feasible to successfully combine the aforementioned approaches, and transform into a  three- and two-level scale both the crowd and the expert judgments.

\subsection{\ref{i:RQ3}: Worker Background and Bias}\label{sec:worker_background_and_bias}

To address \ref{i:RQ3} we study if the answers to questionnaire and CRT test have any relation to worker quality.

\subsubsection{Questionnaire}

\begin{table}[tbp]
    \centering
     \caption{
     Count of the number of workers depending on:  
     number of statements correctly classified (columns, max is 6), vs. (top table) the
     answers to the Political views question (rows) 
     and vs. (bottom table) the number of correct answers to the CRT test (rows, max is 3).
     The last two columns show  the Accuracy (the fraction of correctly identified statements for each group) and the \cem score. 
     }
    \label{tab:q_political_views}
    \scalebox{0.80}{
    \begin{tabular}{r rrrrrrrr r rr r r}
\toprule
& \multicolumn{8}{c}{\textbf{Correctly classified statements}} &&  \textbf{Acc} && \textbf{\cem} \\
&  \textbf{0} &  \textbf{1} & \textbf{2} & \textbf{3} & \textbf{4} & \textbf{5} & \textbf{6} & \textls[-45]{\textbf{Sum}} && \textls[-45] && {\textbf{Mean}}  \\
\cmidrule{2-9}\cmidrule{11-11}\cmidrule{13-13}
\textls[-45]{\textbf{Very conservative}}     &  4 &  3 &   1 &  0 &  0 &  0 &  1 &    9 && .13 && .46 \\
\textbf{Conservative}           & 0 &  9 &   2 &  3 &  1 &  0 &  0 &   15 &&               .21 && .51 \\
\textbf{Moderate}    &  6 &  6 &   6 &  7 &  0 &  1 &  0 &   26 &&                         .20 && .50 \\
\textbf{Liberal}                     &  2 &  8 &  13 &  4 &  4 &  2 &  0 &   33 &&         .16 && .50 \\
\textbf{Very Liberal}                     &  0 &  2 &   6 &  6 &  2 &  1 &  0 &   17 &&    .21 && .51 \\
\cmidrule{2-9}
\textbf{Sum} &  12 &  28 &  28 &  20 &  7 &  4 &  1 &  100 &&&   \\
\end{tabular}}
    \scalebox{0.80}{
    \begin{tabular}{lc rrrrrrrrr r rr r r}
\toprule
&& \multicolumn{8}{c}{\textbf{Correctly classified statements}} && \textbf{Acc} && \textbf{\cem} \\
 &&  \textbf{0} &  \textbf{1} & \textbf{2} & \textbf{3} & \textbf{4} & \textbf{5}& \textbf{6} & \textbf{Sum} && && \multicolumn{1}{c}{\textbf{Mean}} \\
\cmidrule{3-10}\cmidrule{12-12}\cmidrule{14-14}
\textbf{CRT}& \textbf{0}         &  5 &   11 &   9 &  4 &   0 &  1 & 1 & 31 &&.14 && .48 \\
\textbf{correct} & \textbf{1}    &  5 &   10 &   12&  6 &   1 &  0 & 0 & 34 &&.22 && .53 \\
\textbf{answers} & \textbf{2}    &  1 &   6  &   1 &  6 &   3 &  1 & 0 & 18 &&.21 && .51 \\
& \textbf{3}                     &  1 &   1 &    6 &  4 &   3 &  2 & 0 & 17 &&.15 && .47 \\
\cmidrule{3-10}
& \textbf{Sum} &  12 &  28 &  28 &  20 &  7 & 4 &1 &  100 &&&   \\
\bottomrule
\end{tabular}}
\end{table}

Table~\ref{tab:q_political_views} (top) shows in the rows the answer to the workers political views,
while on the columns the number of correctly classified statements (columns, max is 6). 
As we can see from the table, there is only one worker who successfully classified all 6 statements. 
Many workers correctly classified 1 or 2 statements (28 and 28, respectively). The next column summarizes, using Accuracy (i.e., the fraction of exactly classified statements), the quality of workers in each group.
The number and fraction of correctly  classified statements are however rather crude measures of worker's quality, as small misclassification errors (e.g, \politifactpantsfire in place of \politifactfalse) are as important as more striking ones (e.g., \politifactpantsfire in place of \politifacttrue). Therefore, to
 measure the ability of  workers to correctly classify the statements,
we also compute \cem, an effectiveness metric recently proposed  for  the specific case of ordinal classification \cite{ACL:2020} (see \citet[\S3.3]{SIGIR:2020} for a more detailed discussion of these issues). The last column in the table shows the average \cem value for the workers in each group.
By looking at both Accuracy and \cem, it is clear that `Very conservative' workers provide lower quality labels. The Bonferroni corrected two tailed t-test
on \cem confirms that `Very conservative' workers perform statistically significantly worse than both `Conservative' and `Very liberal' workers. 
The workers' political views affect the \cem score, even if in a small way and mainly when considering the extremes of the scale.
An initial analysis of the other answers to the questionnaire (not shown due to space limitations) does not seem to provide strong signals; a more detailed analysis is left for future work. 

\subsubsection{CRT Test}
We now investigate the effect of the CRT test on the worker quality.
Table~\ref{tab:q_political_views} (bottom) shows the count of the number of workers depending on: number of statements correctly classified (columns, max is 6), versus the number of correct answers to the CRT test (rows, max is 3).
Concerning CRT scores, we see that the minority of workers (17) answered in a correct way to all the three questions, and the majority of them  answered correctly to only 1 CRT question (34) or none (31).
Although there is some variation in both Accuracy and \cem, this is never statistically significant; it appears that the number of correct answers to the CRT test is not correlated with worker quality. We leave for future work a more detailed study of this aspect.

\subsection{\ref{i:RQ4}: Worker Behavior}\label{sec:worker_behavior}

We now turn to \ref{i:RQ4}, and analyze the behavior of the workers while performing the task.

\subsubsection{Time}

\begin{figure}[tbp]
  \centering
  \begin{tabular}{@{}c@{}c@{}}
    \includegraphics[width=.55\linewidth]{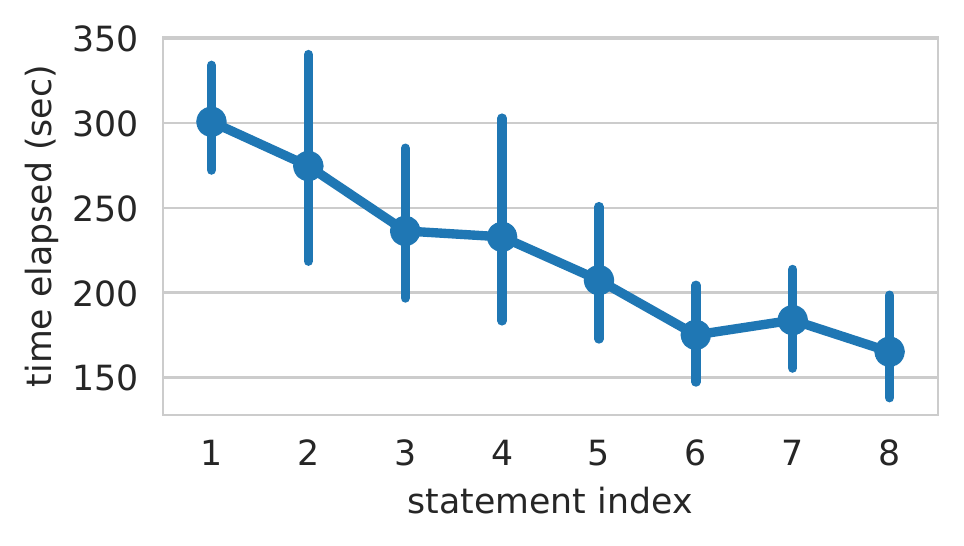}&
    \includegraphics[width=.45\linewidth]{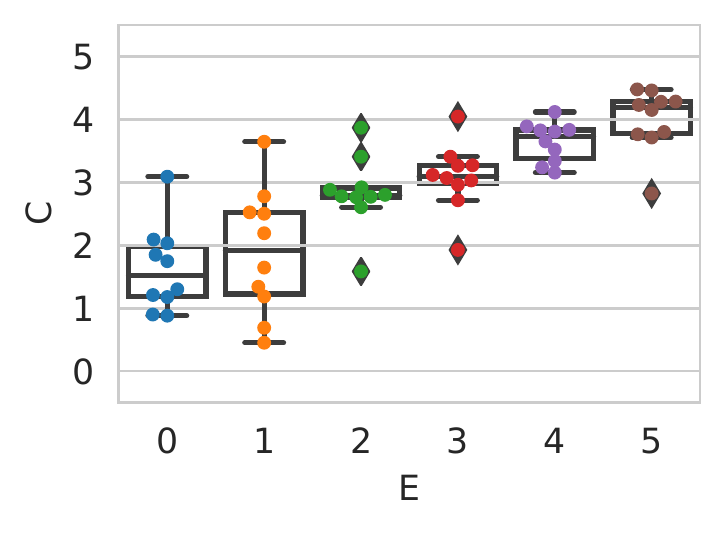}
  \end{tabular}
\caption{%
 Position of the statement in the task vs. 
 time elapsed, cumulative on each single statement
 (left). Comparison between  \expertsix and \six where the aggregation function is the weighted mean and the weights are 
 the political views (see Table~\ref{tab:q_political_views} top) normalized to $[0.5,1]$ (right).
}
  \label{fig:learning_effect_and_political}
\end{figure}

Figure~\ref{fig:learning_effect_and_political} (left) shows that the amount of time spent on average by the workers on the first statements is considerably higher than on the last statements.
This, combined with the fact that the quality of the assessment provided by the workers does not decrease for the last statements
(\cem scores per position are 
1: .61,
2: .60,
3: .64,
4: .58,
5: .59,
6: .54,
7: .61,
8: .62),
is an indication of a learning effect: the workers learn how to assess truthfulness in a faster way.

\subsubsection{Exploiting Worker Signals to Improve Quality}\label{sec:exploiting}

We have shown that, while performing their task, workers provide many signals that to some extent correlate with the quality of their work. These signals could in principle be exploited to aggregate the individual judgments in a more effective way (i.e., giving more weight to workers that possess features indicating a higher quality). For example, the relationships between worker background / bias and worker quality (Section~\ref{sec:worker_background_and_bias}) could be exploited to this aim.

We thus performed the following experiment: we aggregated \six individual scores, using as aggregation function a weighted mean, where the weights are represented by the political views, normalized to $[0.5,1]$. 
Figure~\ref{fig:learning_effect_and_political} (right)  shows the results. 
We also aggregated \six individual scores using as aggregation function the weighted mean function where the weights are represented by the number of correct answers to CRT, normalized to $[0.5,1]$, which lead to similar results. 
Thus, it seems that leveraging quality-related behavioral signals, like questionnaire answers or CRT scores, to aggregate results does not provide a noticeable increase in the external agreement, although it does not harm. We have only scratched the surface, though, as there are many other signals, and aggregation functions, that can be tried; we leave for future work the in depth analysis of how such behavioral signals can be leveraged to improve external agreement.

\subsubsection{Queries}

\begin{table}[tbp]
    \centering
        \caption{
     Statement position in the task versus:
     number of queries issued (top) and 
     number of times the statement has been used as a query (bottom).
     }
    \label{tab:query_stats}
    \scalebox{0.7}{
\begin{tabular}{m{1.5cm}m{0.6cm}m{0.6cm}m{0.6cm}m{0.6cm}m{0.6cm}m{0.6cm}m{0.6cm}m{0.6cm}m{0.6cm}m{0.6cm}}
\toprule
\textbf{Statement Position} & \textbf{1} & \textbf{2} & \textbf{3}  & \textbf{4} & \textbf{5} & \textbf{6} & \textbf{7} & \textbf{8} & \textbf{Sum}& \textbf{Mean}\\
\midrule
\textbf{Number of Queries}& 352 16.8\% & 280 13.4\% & 259 12.4\% & 255 12.1\% & 242 11.6\% & 238 11.3\% &230 11.0\% & 230 11.4\% & 2095 & 261.9\\
\midrule
\textbf{Statement as Query}& 22 9\% & 32 13\% & 31 12.6\% & 33 13.5\%  & 34 13.9\% & 30 12.2\% & 29 11.9\% & 34 13.9\% & 245 & 30.6\\
\bottomrule
\end{tabular}}
\end{table}

Table~\ref{tab:query_stats} shows query statistics for the 100 workers which finished the task. 
As we can see, the higher the statement position, the lower the number of queries issued: 3.52 queries on average for the first statement, down to 2.3 for the last statement. This can indicate the attitude of workers to issue fewer queries the more time they spend on the task, probably due to fatigue, boredom, or learning effects. 
Nevertheless, we can see that on average, for all the statement positions each worker issues more than one query, i.e., workers often reformulate their initial query. This provides further evidence that they put effort in performing the task. 
The third row of the table shows the number of times the worker used as query the whole statement. 
We can see that the percentage is rather low (around 13\%) for all the statement positions, indicating again that workers spend effort when providing their judgments.

\subsection{\ref{i:RQ5}: Sources of Information}\label{sources_of_information}

\subsubsection{URL Analysis}\label{sec:url_analysis}

\begin{figure}[t]
  \centering
  \begin{tabular}{@{}c@{}}
    \includegraphics[width=.45\linewidth]{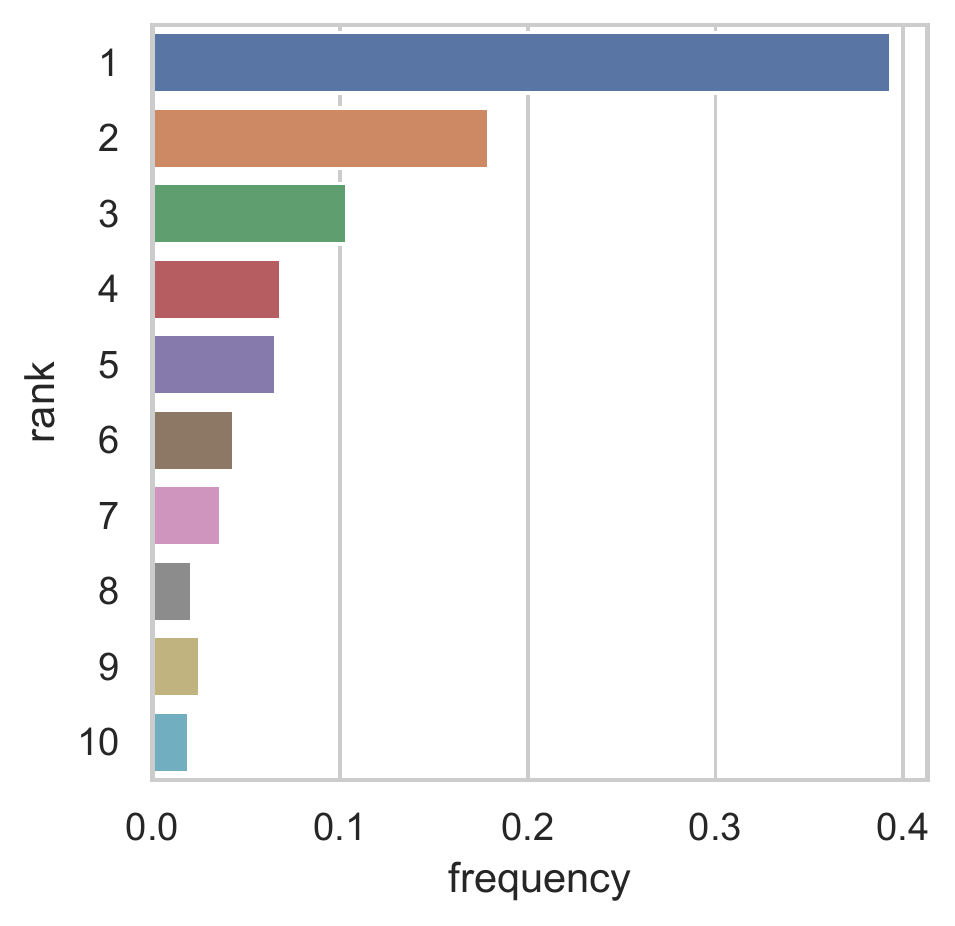} 
  \end{tabular}
  {\small
      \scalebox{0.8}{
  \begin{tabular}{l@{~}r@{\%}}
  \toprule
        \textbf{URL} &          \textbf{Percentage}\\ 
\midrule
              snopes.com &            11.79 \\
                 msn.com &             8.93 \\
           factcheck.org &             6.79 \\
                wral.com &             6.79 \\
            usatoday.com &             5.36 \\
           statesman.com &             4.64 \\
             reuters.com &             4.64 \\
                 cdc.gov &             4.29 \\
  mediabiasfactcheck.com &             4.29 \\
     businessinsider.com &             3.93 \\
\bottomrule
  \end{tabular}}
  }
\caption{%
    On the left, distribution of the ranks of the URLs selected by workers,
    on the right, websites from which workers chose URLs to justify their judgments.
  \label{fig:url-ranks-distributions}}
\end{figure}

Figure~\ref{fig:url-ranks-distributions} shows on the left the distribution of the ranks of the URL selected as evidence by the worker when performing each judgment. 
URLs selected less than 1\% times are filtered out from the results. 
As we can see from the plot, about 40\% of workers selected the first result retrieved by our search engine, and selected the remaining positions less frequent, with an almost monotonic decreasing frequency (rank 8 makes the exception). We also found that 14\% of workers inspected up to the fourth page of results (i.e., rank$=40$). The breakdown on the truthfulness \politifact categories does not show any significant difference.

Figure~\ref{fig:url-ranks-distributions} shows on the right part the top 10 of websites from which the workers choose the URL to justify their judgments. Websites with percentage $\leq 3.9\%$ are filtered out. As we can see from the table, there are many fact check websites among the top 10 URLs (e.g., snopes: 11.79\%, factcheck 6.79\%).
Furthermore, medical websites are present, although in small percentage (cdc: 4.29\%). 
This indicates that workers use various kind of sources as URLs from which they take information. Thus, it appears that they put effort in finding evidence to provide a reliable truthfulness judgment.

\subsubsection{Justifications}\label{sect:justifications_behavior}
As a final result, we analyze the textual justifications provided, their relations with the web pages at the selected URLs, and their links with worker quality.
54\% of the provided justifications contain text copied from the web page at the URL selected for evidence, while 46\% do not. 
Furthermore, 48\% of the justification include some ``free text'' (i.e., text generated and written by the worker), and 52\% do not.
Considering all the possible combinations,
6\% of the justifications used both free text and text from web page,
42\% used free text but no text from the web page,
48\% used no free text but only text from web page, and finally
4\% used neither free text nor text from web page, and either inserted text from a page of a different (not selected) web page or inserted part of the instructions we provided or text from the user interface.

Concerning the preferred way to provide justifications, each worker seems to have a clear attitude:
48\% of the workers used only text copied from the selected web pages,
46\% of the workers used only free text,
4\% used both,
and 2\% of them consistently provided text coming from the user interface or random web pages.

\begin{figure}[tbp]
  \centering
  \begin{tabular}{@{}c@{}c@{}}
     \includegraphics[width=.49\linewidth]{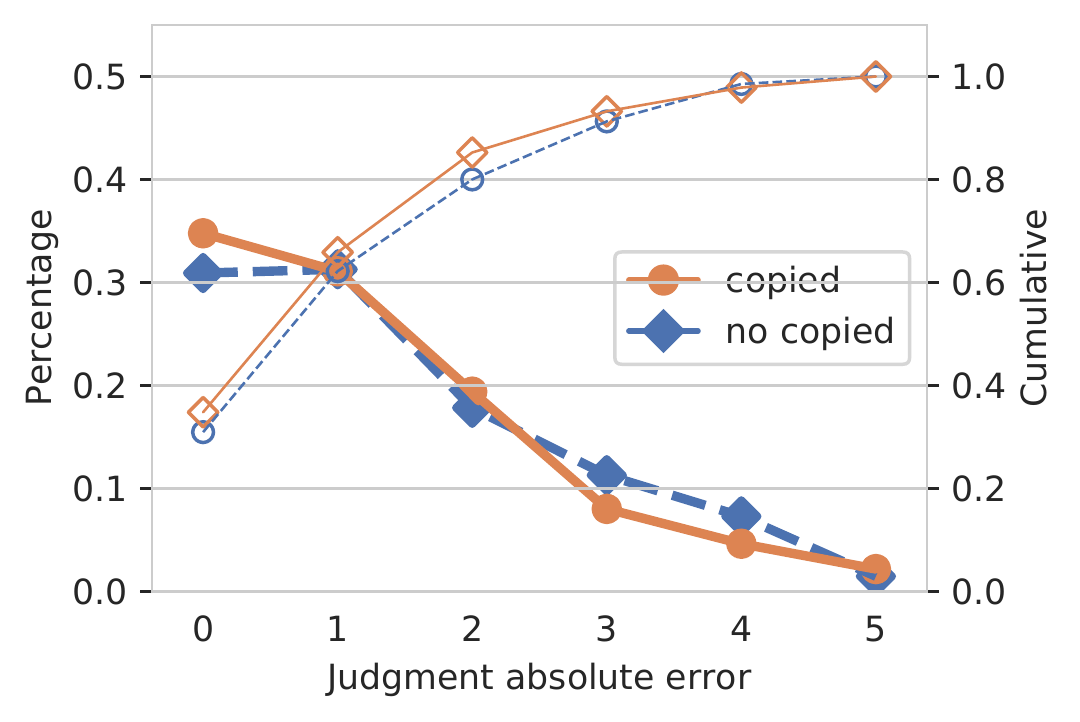}&
     \includegraphics[width=.5\linewidth]{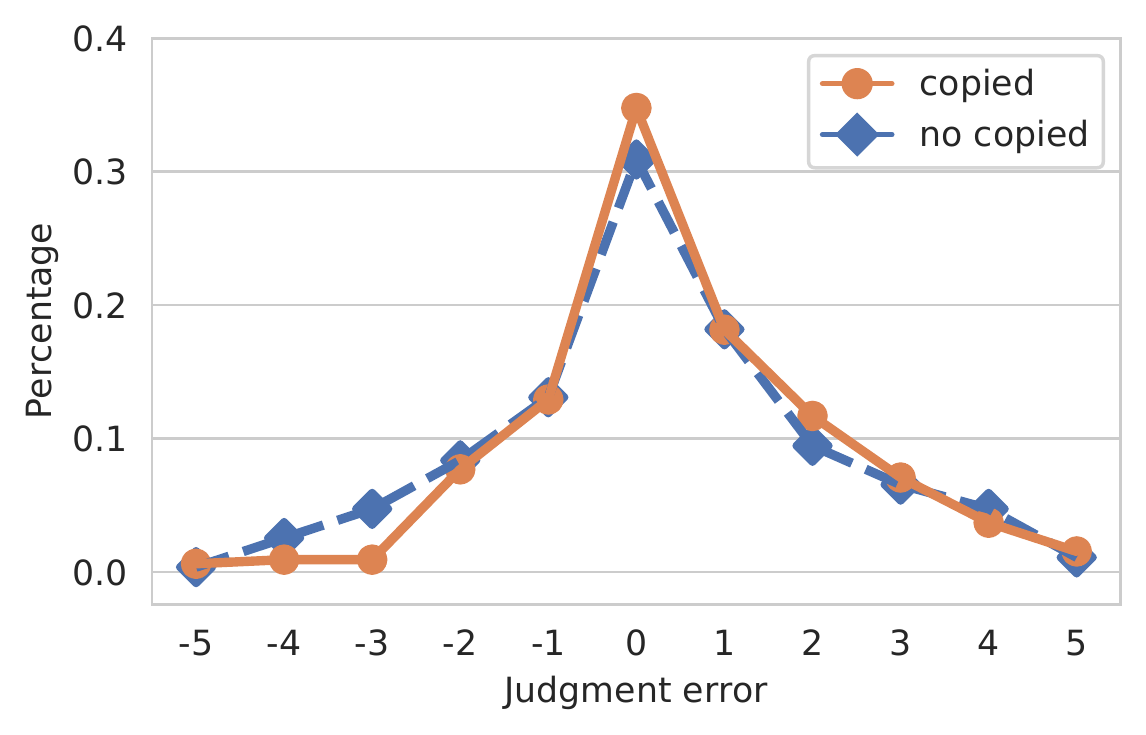} \\
  \end{tabular}
\caption{%
    Effect of the origin of a justification (text copied/not copied from the URL selected) on: 
    the absolute value of the prediction error (left; cumulative distributions shown with thinner lines and empty markers), and 
    the prediction error (right).
    }
  \label{fig:justification_error}
\end{figure}

We now correlate such a behavior with the workers quality.
Figure~\ref{fig:justification_error} 
shows the relations between different kinds of justifications and the worker accuracy.
The plots show the absolute value of the prediction error on the left, and the prediction error on the right. 
The lines in the plots indicate if the text inserted by the worker was copied or not from the web page selected.
We did the same analysis to investigate if the worker used or not free text, and the plots where almost indistinguishable. 

As we can see from the plot 
, statements on which workers make less errors (i.e., where x-axis$=0$) tend to use text copied from the web page selected. 
On the contrary, statements on which workers make more errors (values close to 5 in the left plot, and values close to +/- 5 in the right plot) tend to use text not copied from the selected web page. 
The differences are small, but it might be an indication that workers of higher quality tend to read the text from selected web page, and report it in the justification box.
To confirm this result, we computed the \cem scores for the two classes considering the individual judgments:
the class ``copied'' has \cem$=0.62$, while
the class ``not copied'' has a lower value, \cem$=0.58$.
The behavior is consistent for what concerns the usage of free text (not shown).

By looking at the right column of Figure~\ref{fig:justification_error} we can see that the distribution of the prediction error is not symmetrical, as the frequency of the errors is higher on the positive side of the x-axis ([0,5]). These errors correspond to  workers overestimating the truthfulness value of the statement (with 5 being the result of labeling a \politifactpantsfire statement as \politifacttrue). This is consistent with what observed in Sect.~\ref{sec:crowd_score_distribution}.
It is also noticeable that the justifications containing text copied from the selected URL have a lower rate of errors in the negative range, meaning that workers which directly quote the text avoid underestimating the truthfulness of the statement.
These could be other useful signals to be exploited in future work to obtain more effective aggregation methods.

\section{Conclusions and Future Work}\label{sec:concl}

The work presented in this paper is, to the best of our knowledge, the first one investigating the ability of crowd workers to identify and correctly categorize recent health statements related to the \covid pandemic.
The workers performed a task consisting of judging the truthfulness of 8 statements using our customized search engine, which allows us to control worker behavior. 
We analyze workers background and bias, as well as workers cognitive abilities, and we correlate such information to the worker quality.
We publicly release the collected data to the research community.

The answers to our research questions can be summarized as follows.
We found evidence that the workers are able to detect and objectively categorize online (mis)information related to the \covid pandemic (\ref{i:RQ1}).
We found that while the agreement among workers does not provide a strong signal, aggregated workers judgments show high levels of  agreement with the expert labels, with the only exception of the two truthfulness categories at the lower end of the scale (\politifactpantsfire and \politifactfalse).
We found that both crowdsourced and expert judgments can be transformed and aggregated to improve label quality (\ref{i:RQ2}).
We found that worker political background, self-reported in a questionnaire, is indicative of label quality 
(\ref{i:RQ3}). 
We found several promising behavioral signals that are clearly related with worker quality (\ref{i:RQ4}). Such signals may effectively inform new ways of aggregating crowd judgments (e.g., see \cite{raykar2010learning,baba2013statistical}), which we believe is a promising direction for future  work.
Finally, 
we found that workers use multiple sources of information, and they consider both fact-checking and health-related websites. We also found interesting relations between the justifications provided by the workers and the judgment quality (\ref{i:RQ5}).
Future work also includes reproducing our experiments in other crowdsourcing platforms to target other cohorts of workers.

\begin{acks}
This work is partially supported by 
a \grantsponsor{}{Facebook Research}{https://research.fb.com/programs/research-awards/proposals/the-online-safety-benchmark-request-for-proposals/} award,
by an \grantsponsor{DP190102141}{Australian Research Council
Discovery Project (DP190102141)}{http://purl.org/au-research/grants/arc/DP190102141}, 
and 
by the project \grantsponsor{1619942002 / 1420AFPLO1}{HEaD – Higher Education and Development - 1619942002 / 1420AFPLO1 (Region Friuli – Venezia Giulia)}{http://www.innovafvg.it/uploads/media/DRN_788_2019_Bando_4_assegniHEaD-S3.pdf}.
\end{acks}

\bibliographystyle{ACM-Reference-Format}
\bibliography{zzz-reference}

\end{document}